\documentclass[11pt]{article}
\usepackage[journal=jquant,stage=submission,logging=stdout]{jsty3-author}
\usepackage{tikz}
\usetikzlibrary{arrows.meta}

\DraftwatermarkOptions{stamp=false}
\AtBeginDocument{\nolinenumbers}

\ExplSyntaxOn
\cs_set_protected:Npn \__jsty_print_logo: {}
\ExplSyntaxOff

\newsavebox{\RSContourBox}
\newif\ifRSContoursReady
\newcommand{\RSInsertContours}{%
 \par\noindent\usebox{\RSContourBox}\par\medskip
 \global\RSContoursReadyfalse
}
\newcommand{\PlaceRSContoursOnPageFour}{%
 \global\RSContoursReadytrue
 \ifnum\value{page}>3\relax\RSInsertContours\fi
}
\AddToHook{cmd/@outputpage/after}[RScontours]{%
 \ifnum\value{page}=4\relax
  \ifRSContoursReady\RSInsertContours\fi
 \fi
}

\title{A Quantum Model from the Riemann–Siegel Formula}
\author{a,b}{Germ\'an Sierra}{}{}
\affiliation{a}{Instituto de F\'{\i}sica Te\'orica, CSIC-UAM, Madrid, Spain}
\affiliation{b}{Quantum Advanced Center Research, Madrid, Spain}
\abstract{We construct a two-component quantum model in which
destructive interference selects the critical-line zeros of a
regularized zeta function within a continuous spectrum. Starting
from the exact Riemann--Siegel integral representation, we define
$\zeta_{\rm reg}(s)$ by subtracting the contribution of the pole at
the origin. The two remaining contour integrals supply the
interfering amplitudes.

A Dirichlet heat filter suppresses high spatial frequencies, and an
arithmetic dressing has a profile generated by prime-power
dilations. These transformations map the dilation generator to a
self-adjoint, parity--time ($PT$) symmetric Hamiltonian on a physical
Hilbert space with a positive metric. Its spectrum is real and
continuous. The computed lower critical-line zeros of
$\zeta_{\rm reg}(s)$ lie close to the corresponding Riemann zeros and
exhibit spacing and pair-correlation statistics consistent with the
Gaussian unitary ensemble over the sampled range. Whether every
analytic zero in the lower half of the critical strip belongs to
the physical spectral family remains open.}

\AddToHook{cmd/tableofcontents/before}{%
  \begingroup
  \setcounter{page}{2}%
  \fontsize{10.5}{13}\selectfont
  \setlength{\cftbeforesecskip}{2pt}%
  \setlength{\cftbeforesubsecskip}{0pt}%
  \setlength{\cftbeforesubsubsecskip}{0pt}%
}
\AddToHook{cmd/tableofcontents/after}{\endgroup}

\begin{document}

\clearpage
\section{Introduction}
The Riemann hypothesis (RH) states that all nontrivial zeros of the
Riemann zeta function lie on the critical line $\Re s=1/2$
\cite{Riemann1859,BombieriRH,SarnakRH,ConreyRH,Titchmarsh86,Edwards,BorweinEtAl08}.
The Hilbert--P\'olya proposal seeks a self-adjoint operator whose
spectrum encodes their ordinates \cite{Polya-Hilbert}. The connection
between zero statistics, random matrices, and quantum chaos
\cite{Montgomery73,Odlyzko87,Berry86} has motivated spectral proposals
ranging from the Berry--Keating dilation model to Connes's interpretation
of the zeros as an absorption spectrum
\cite{BerryKeatingXP99,BK99,Connes99}. For a review of physical approaches
to RH, see Ref.~\cite{SchumayerHutchinson2011}.

Here we approach the
spectral question through the Riemann--Siegel (RS) integral
representation of the Riemann zeta function \cite{Siegel1932,SiegelTranslation18}.
The RS  formula has served two distinct purposes in earlier
spectral works. It guided the semiclassical resummation of periodic-orbit
contributions to spectral determinants \cite{BerryKeating1990,Keating1992}
and was used to construct boundary wave functions in an interacting
$xp$ model \cite{Sierra2008RiemannZeros}. A separate motivation of the present work comes
from Dirac models in Rindler spacetime. The massive model reproduces
the smooth part of the zero-counting function, while the massless model
with delta-function scatterers explores how arithmetic information can
enter the spectrum \cite{GS16}.

Our construction starts from the exact integral representation
recovered by Siegel from Riemann's unpublished notes. After separating the pole
contribution, we use the two regular integrals to construct a spinor.
Zeros of the regularized function occur where the two integrated
amplitudes cancel. The resulting differential Hamiltonian combines a
dilation term with a non-Hermitian second-order term and has an internal
parity--time ($PT$) symmetry. A Dirichlet heat transformation relates it
to a self-adjoint dilation generator. Transporting the Hilbert-space
product yields a positive metric and a real continuous spectrum, whose
integral-family states have generalized, rather than finite-norm,
normalization. The cancellation condition selects points of this
continuum. Its exhaustive correspondence with the analytic zeros and
a discrete operator realization remain open.

We prove that the regularized function has no zeros in the upper half
of the critical strip. In the lower half, we numerically locate zeros
on the critical line close to the Riemann zeros and find that their
unfolded spacings and pair correlations are consistent with Gaussian
unitary ensemble statistics. We formulate the regularized Riemann
hypothesis (RRH), which asserts that every zero of the regularized
function in the lower half of the critical strip lies on the critical
line, and conjecture its equivalence with RH.

\section{Riemann--Siegel formula and regularization}
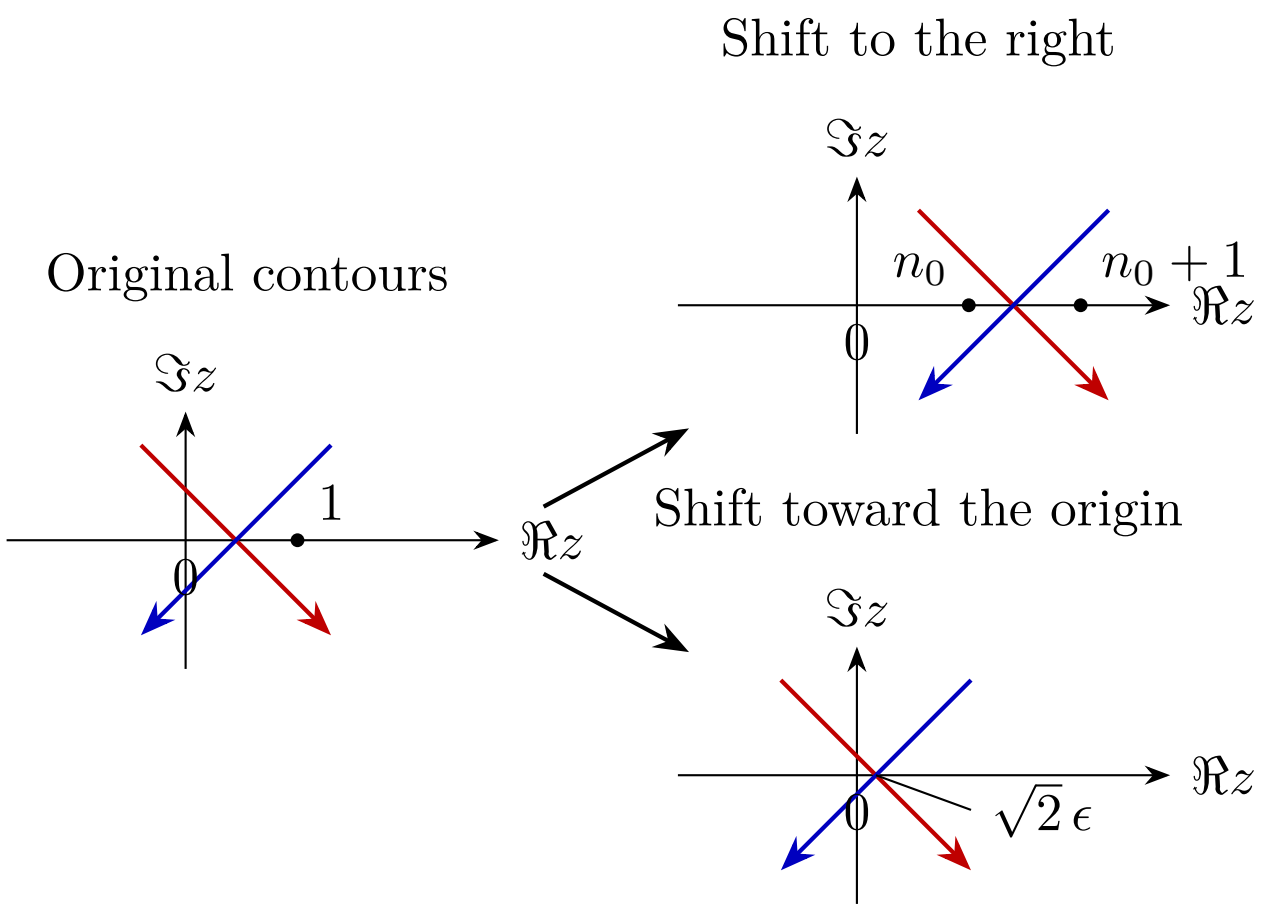
\begin{figure}[t]
 \centering
 \begin{tikzpicture}[x=0.74cm,y=0.74cm,>=Stealth,font=\small]
 \foreach \x/\y/\v/\heading in {0/0/.45/{Original contours},
   6/2.1/1.40/{Shift to the right},6/-2.1/.17/{Shift toward the origin}} {
   \begin{scope}[shift={(\x,\y)}]
    \draw[->] (-1.6,0)--(2.8,0) node[right] {$\Re z$};
    \draw[->] (0,-1.15)--(0,1.15) node[above] {$\Im z$};
    \draw[red!75!black,thick,->] (\v-.85,.85)--(\v+.85,-.85);
    \draw[blue!75!black,thick,<-] (\v-.85,-.85)--(\v+.85,.85);
    \node[anchor=east,below] at (0,0) {$0$};
    \node[above=17pt] at (.55,1.15) {\heading};
   \end{scope}
 }
 \fill (1,0) circle (1.3pt) node[above right] {$1$};
 \fill (7,2.1) circle (1.3pt) node[above left] {$n_0$};
 \fill (8,2.1) circle (1.3pt) node[above right] {$n_0+1$};
 \draw (6.17,-2.1) -- (7.02,-2.41) node[right] {$\sqrt2\,\epsilon$};
 \draw[->,thick] (3.2,.3)--(4.5,1.0);
 \draw[->,thick] (3.2,-.3)--(4.5,-1.0);
 \end{tikzpicture}

 \caption{Parallel deformations of the oriented Riemann--Siegel
 contours in the complex $z$ plane. The original red and blue contours
 (left) have slopes $-1$ and $+1$, respectively, and intersect the real
 axis between $0$ and $1$. The upper-right panel shows the deformation
 across the poles at $1,\ldots,n_0$, with the new crossing between
 $n_0$ and $n_0+1$, where $n_0=\lfloor\sqrt{t/(2\pi)}\rfloor$ for $t>0$.
 The lower-right panel shows the deformation used here, with the
 real-axis crossing at $\sqrt{2}\,\epsilon$ and
 $\epsilon\to0^+$. The colored arrowheads indicate the integration
 orientations in Eq.~\eqref{3.6}.}
 \label{fig:RS-contours}
\end{figure}

For the completed zeta function
$\Lambda(s)=\pi^{-s/2}\Gamma(s/2)\zeta(s)$, the exact integral formula is
\begin{align}
\Lambda(s)={}&\pi^{-s/2}\Gamma(s/2)
\int_{0\searrow1}dz\,\frac{z^{-s}e^{-i\pi z^2}}{2i\sin(\pi z)}
\nonumber\\
&+\pi^{(s-1)/2}\Gamma((1-s)/2)
\int_{0\swarrow1}dz\,\frac{z^{s-1}e^{i\pi z^2}}{2i\sin(\pi z)}.
\label{3.6}
\end{align}
The contours are straight lines with slopes $-1$ and $+1$, respectively, crossing the real axis between zero and one.
The oriented contours are shown in figure~\ref{fig:RS-contours}.

 For $s=\tfrac12+it$, with $t \in \mathbb{R}$, shifting these contours to the right across the poles at $n=1,\ldots,n_0$, where $n_0=\lfloor\sqrt{t/(2\pi)}\rfloor$, yields the Riemann--Siegel sum representation of $\zeta(\tfrac12+it)$, including its remainder. Here we move the contours towards the origin while keeping $s$ general, without imposing $\Re s=\tfrac12$. To control this limit, we separate the contribution of the pole at the origin from the regular remainder,
\begin{equation}
\csc(\pi z)=\frac1{\pi z}
 +\left[\csc(\pi z)-\frac1{\pi z}\right].
\label{eq:cosecant-subtraction}
\end{equation}
With the boundary prescription of
Appendix~\ref{app:distributional-limit}, this gives
\begin{align}
\Lambda&=\Lambda_{\rm reg}+\Lambda_{\rm sing},\label{4.19}\\
\Lambda_{\rm sing}(s)&=-\frac{e^{-i\pi s/4}}s
 -\frac{e^{-i\pi(s-1)/4}}{1-s}.\label{4.21}
\end{align}
Equivalently,
\begin{equation}
\zeta=\zeta_{\rm reg}+\Delta,\qquad
\begin{cases}
\zeta_{\rm reg}=\pi^{s/2}\Lambda_{\rm reg}/\Gamma(s/2),\\
\Delta=\pi^{s/2}\Lambda_{\rm sing}/\Gamma(s/2).
\end{cases}
\label{eq:zeta-reg-Delta-decomposition}
\end{equation}
Thus the subtraction follows directly from the Riemann--Siegel
integrands, rather than from an independently chosen correction to
$\zeta$. The resulting regular integrals are given below; their
derivation and further properties are collected in
Appendices~\ref{app:distributional-limit}, 
\ref{app:regularized-properties}, and \ref{app:upper-strip-zero-free}

The completed regularized function $\Lambda_{\rm reg}$ is entire:
the pole terms in $\Lambda$ are canceled by those in
$\Lambda_{\rm sing}$, and their remaining quotients have removable
singularities. Consequently, $\zeta_{\rm reg}(s)=
\pi^{s/2}\Lambda_{\rm reg}(s)/\Gamma(s/2)$ is entire. In particular,
$\zeta_{\rm reg}$ has no pole at $s=1$, unlike $\zeta$, and the zeros of
the reciprocal gamma factor imply
\begin{equation}
 \zeta_{\rm reg}(-2n)=0,\qquad n=0,1,2,\ldots.
 \label{eq:main-regularized-trivial-zeros}
\end{equation}
For $n\geq1$ these include the trivial zeros of $\zeta$; the zero at
$s=0$ is an additional zero of $\zeta_{\rm reg}$.

Two properties distinguish the regularized zero problem.
 First, it obeys a critical-line reflection symmetry,
\begin{equation}
 \Lambda_{\rm reg}(s)=\overline{\Lambda_{\rm reg}(1-\bar s)},
 \label{eq:main-cross-reflection}
\end{equation}
so off-line zeros occur in pairs $\{s,1-\bar s\}$ at the same
imaginary ordinate. Second, the upper critical strip is zero-free:
\begin{equation}
 \begin{aligned}
 \Re\Lambda_{\rm reg}(\sigma+it)
 &\leq\frac{2}{\pi(e^\pi-1)}-\left(1-\frac1{\sqrt2}\right) < -0.264, \\
 & 0<\sigma<1,\quad t\geq0.
 \end{aligned}
 \label{eq:upper-strip-bound}
\end{equation}
Since the gamma prefactor is nonzero in the strip, this implies
\begin{equation}
 \zeta_{\rm reg}(s)\ne0,
 \qquad 0<\Re s<1,\quad\Im s\geq0.
 \label{eq:upper-strip-zero-free}
\end{equation}
The proof is deferred to Appendix~\ref{app:upper-strip-zero-free}.
All critical-strip regularized zeros therefore lie in the lower
half-strip. For fixed $0<\sigma<1$, their relation to the original
function is asymptotically controlled by
$\Delta(\sigma-iT)=O(T^{-(1+\sigma)/2})$ as $T\to\infty$.

We formulate the \emph{regularized Riemann hypothesis} (RRH) as
\begin{equation}
 \left.
 \begin{gathered}
 \zeta_{\rm reg}(s)=0,\\[-2pt]
 0<\Re s<1
 \end{gathered}\right\}
 \Longrightarrow\Re s=\tfrac12.
 \label{eq:regularized-RH}
\end{equation}
By Eq.~\eqref{eq:upper-strip-zero-free}, RRH places all critical-strip
zeros of $\zeta_{\rm reg}$ on the lower critical line. We conjecture
that RRH is equivalent to RH; this global equivalence remains open.

A local comparison follows by linearizing $\Lambda(s_0+h)=0$ at a
regularized zero $s_0$ on the critical line ($\Lambda_{\rm reg}(s_0) =0$):
\begin{equation}
 h_{\rm lin}=-\frac{\Lambda_{\rm sing}(s_0)}{\Lambda'(s_0)}
 \in i\mathbb R,
 \label{eq:main-local-zero-shift}
\end{equation}
provided $\Lambda'(s_0)\ne0$. The estimate is tangent to the critical
line; its accuracy requires control of the Taylor remainder and the
existence of a nearby zero. Appendix~\ref{app:perturbative-zero-correspondence}
discusses both directions and proves a local correspondence for simple
zeros under sufficiently small real perturbations.

For a numerical comparison, write
\begin{equation*}
 \zeta(\tfrac12+i\gamma_n)=0,\qquad
 \zeta_{\rm reg}(\tfrac12-i\widetilde\gamma_n)=0,
\end{equation*}
with both positive sequences ordered increasingly.
Table~\ref{tab:zero-ordinates} compares the first ten ordinates, their
differences, and the linear estimate $h_n=i h_{\rm lin}$ of the
ordinate shift. The latter is evaluated at
$s_0=\tfrac12-i\widetilde\gamma_n$ using
Eq.~\eqref{eq:main-local-zero-shift}. The observed
agreement concerns this finite sample and does not establish a global
pairing of zeros.

\begin{table}[t]
\centering
\begin{tabular}{rrrr}
\hline
$\gamma_n$ & $\tilde{\gamma}_n$ & $\gamma_n-\tilde{\gamma}_n$ & $h_n$ \\
\hline
14.1347 & 14.0701 & 0.0647 & 0.0618 \\
21.0220 & 21.0544 & -0.0324 & -0.0333 \\
25.0109 & 24.9874 & 0.0234 & 0.0231 \\
30.4249 & 30.4460 & -0.0211 & -0.0216 \\
32.9351 & 32.9163 & 0.0187 & 0.0186 \\
37.5862 & 37.5982 & -0.0120 & -0.0121 \\
40.9187 & 40.9041 & 0.0146 & 0.0144 \\
43.3271 & 43.3384 & -0.0113 & -0.0114 \\
48.0052 & 47.9930 & 0.0122 & 0.0120 \\
49.7738 & 49.7869 & -0.0131 & -0.0132 \\
\hline
\end{tabular}
\caption{Comparison of the first ten positive Riemann ordinates $\gamma_n$ with the positive magnitudes $\tilde{\gamma}_n$ of the negative ordinates of the regularized critical-line zeros. Rows are ordered by increasing ordinate. The third column gives $\gamma_n-\tilde{\gamma}_n$, and the fourth column gives the ordinate-shift estimate $h_n=i h_{\rm lin}$ obtained from Eq.~\eqref{eq:main-local-zero-shift}. Values are rounded to four decimal places.}
\label{tab:zero-ordinates}
\end{table}

To investigate the local statistics, we use the first $10^4$ positive
Riemann ordinates from Odlyzko's tables~\cite{OdlyzkoTables}, refining
them numerically. For each ordinate we use the reverse linear estimate
in Eq.~\eqref{eq:app-reverse-shift} as the initial guess for a numerical
solution of $\zeta_{\rm reg}(\tfrac12-i\widetilde\gamma)=0$ on the
critical line. Both sequences are unfolded using the smooth
Riemann--von Mangoldt counting function.

\begin{figure}[t]
 \centering
 \begin{subfigure}[t]{0.485\textwidth}
  \centering
  \includegraphics[width=\linewidth]{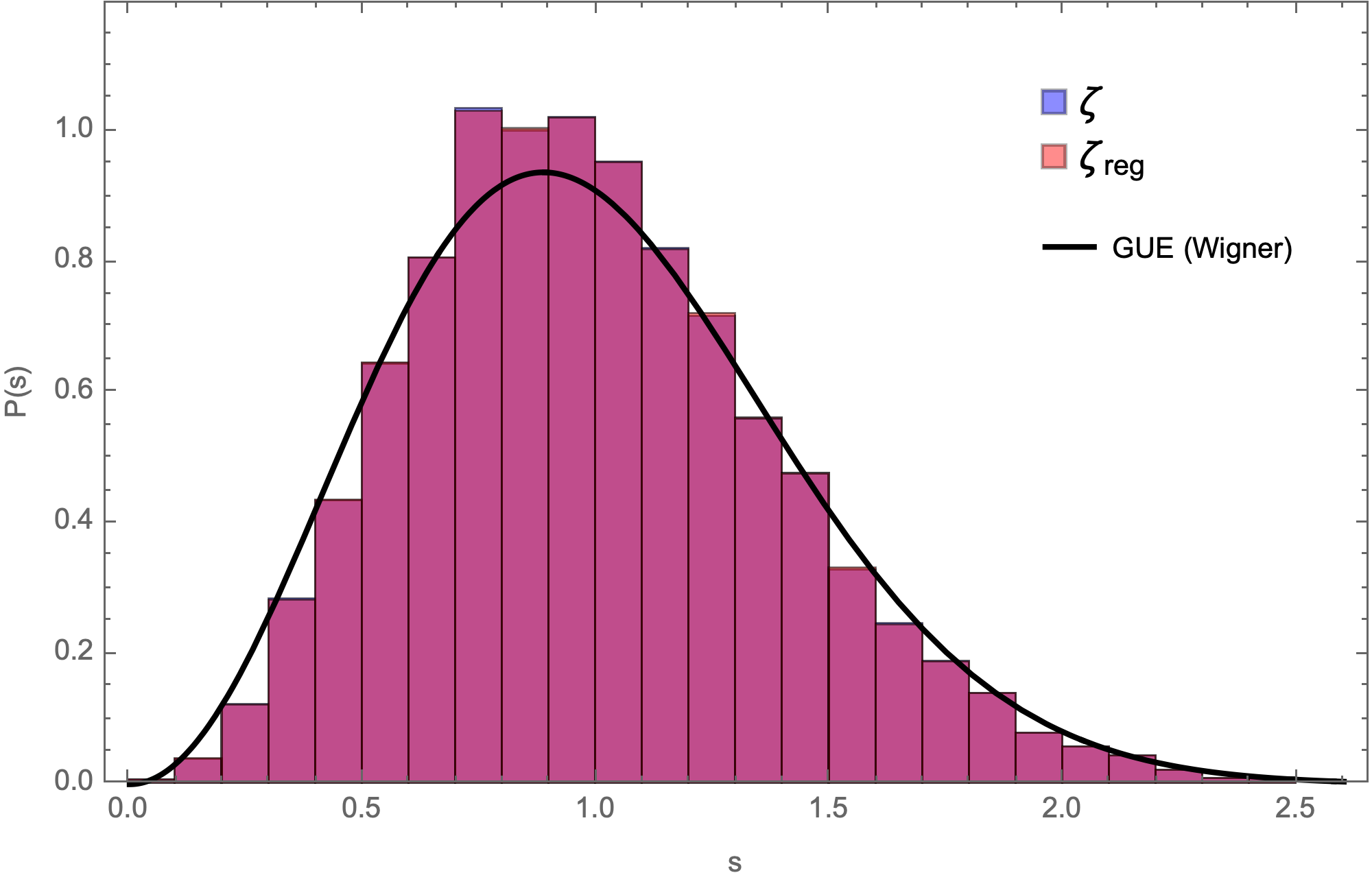}
  \caption{}\label{fig:spacing-gue}
 \end{subfigure}\hfill
 \begin{subfigure}[t]{0.485\textwidth}
  \centering
  \includegraphics[width=\linewidth]{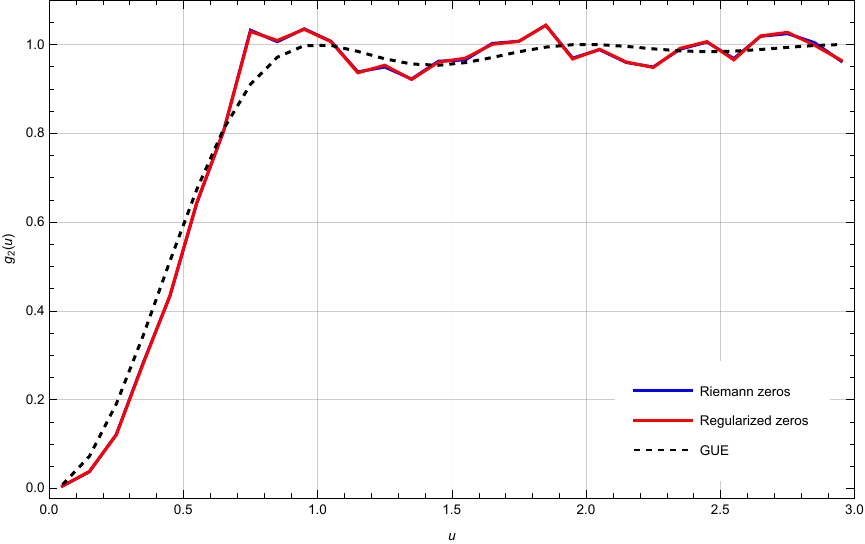}
  \caption{}\label{fig:pair-correlation-gue}
 \end{subfigure}
 \caption{Zero statistics for $10^4$ unfolded Riemann zeros (blue) and
 numerically determined regularized zeros (red). (a) Nearest-neighbor
 spacing distributions; the black curve is the GUE Wigner surmise
 $P_{\rm W}(s)=32s^2e^{-4s^2/\pi}/\pi^2$.
 (b) Pair correlations of the same sequences; the dashed black curve
 is the Montgomery--GUE prediction $g_2(u)=1-[\sin(\pi u)/(\pi u)]^2$.
 Both sequences are unfolded using the smooth Riemann--von Mangoldt
 counting function.}
 \label{fig:gue-statistics}
\end{figure}

Figure~\ref{fig:gue-statistics}(\subref{fig:spacing-gue}) compares the resulting nearest-neighbor
spacing distributions with the GUE Wigner surmise,
$P_{\rm W}(s)=\frac{32}{\pi^2}s^2 e^{-4s^2/\pi}$, an approximation to
the GUE spacing distribution. Figure~\ref{fig:gue-statistics}(\subref{fig:pair-correlation-gue})
compares the pair correlations with the Montgomery--GUE prediction
$g_2(u)=1-[\sin(\pi u)/(\pi u)]^2$~\cite{Montgomery73,Odlyzko87}.
The two numerical sequences show closely matching statistics, broadly
consistent with these GUE benchmarks at the sampled heights. This finite
sample neither proves the asymptotic GUE conjectures nor establishes a
complete correspondence between the two sets of zeroes. 

\section{From contour integrals to spinors}
Our next objective is to construct a quantum model starting from the
regularized zeta function $\zeta_{\rm reg}(s)$. To this end, we recast
its two integral contributions as components of a spinor. This
representation will provide the states used to formulate the
Hamiltonian description in the next section.

Introduce a rescaled regular function
\begin{equation}
R(s)=-\frac{e^{i\pi(s-1/2)/4}}{\sqrt\pi}
 \Gamma\!\left(\frac{1+s}{2}\right)
 \Gamma\!\left(1-\frac{s}{2}\right)\Lambda_{\rm reg}(s).
\label{5.3}
\end{equation}

Throughout $0<\Re s<1$, its prefactor and the completed-zeta
prefactor are nonzero; thus $R$ and $\zeta_{\rm reg}$ have precisely the
same zeros. Set $\alpha_\pm=e^{\mp i\pi/4}$ and
\begin{equation}
 c_\pm(x)=-\pi\alpha_\pm\csc(\pi\alpha_\pm x)+\frac1x.
\label{5.5}
\end{equation}
Gaussian Mellin representations of the gamma factors give
\begin{align}
R(s)={}&2\int_0^\infty\!dp\int_0^\infty\!dx\,e^{-\pi(p^2+x^2)}\label{5.6}\\
&\times\bigl[e^{-i\pi/8}(p/x)^s c_+(x) +e^{i\pi/8}(p/x)^{1-s}c_-(x)\bigr].
\nonumber
\end{align}
All dependence on the spectral parameter $s$
enters through the ratio $p/x$, motivating the following change of variables:
\begin{equation}
 u=p/x,\quad v=px,\quad
 p=\sqrt{uv},\quad x=\sqrt{v/u} .
\label{5.7}
\end{equation}
The Jacobian is $dp\,dx=du\,dv/(2u)$, and the Gaussian factor becomes $K(u,v)=e^{-\pi v(u+u^{-1})}$. Explicitly,
\begin{align}
 R(s)={}&\int_0^\infty\!du\int_0^\infty\!dv\,K(u,v)
 \label{eq:uv-amplitude} \\
 &\times\bigl[e^{-i\pi/8}u^{s-1}c_+(\sqrt{v/u})
+e^{i\pi/8}u^{-s}c_-(\sqrt{v/u})\bigr].
\nonumber
\end{align}
The spectral parameter now enters through conjugate Mellin powers  $s$ and $1-s$.
 The inversion-symmetric combination $u+u^{-1}$ defines a common
damping kernel for both components, while the factors $c_\pm$
encode the contour dependence of the regularized denominators in the Riemann--Siegel integrals.

These odd functions
have the sine Fourier representation

\begin{equation}
 c_\pm(x)=\int_0^\infty dr\,\sin(xr)\widehat c_\pm(r), \quad
 \widehat c_\pm(r)=\frac{2}{1+e^{\alpha_\pm r}}.
\label{5.9}
\end{equation}
For $\Re\alpha>0$, expansion of $(1+e^{\alpha r})^{-1}$ and
elementary sine transforms give
\[
 \int_0^\infty\frac{2\sin(xr)}{1+e^{\alpha r}}\,dr
 =2\sum_{n\ge1}\frac{(-1)^{n-1}x}{n^2\alpha^2+x^2}
 =\frac1x-\frac{\pi}{\alpha\sinh(\pi x/\alpha)}.
\]
Termwise integration follows from $|\sin(xr)|\le |x|r$ near
$r=0$ and exponential decay at infinity. The alternating
partial-fraction expansion of the hyperbolic cosecant gives the
last equality, with removable values understood by continuity.
For $\alpha=\alpha_\pm$ this recovers Eq.~\eqref{5.5}.
The Gaussian factors and the integrable powers $x^{-\Re s}$ and
$x^{\Re s-1}$ at the origin justify exchanging integrals in
$0<\Re s<1$, since the sine weights decay exponentially in $r$.

Define the spinors
\begin{align}
 \Psi_{+/-}(r;s) &= \widehat c_{+/-}(r)I_{s/1-s}(r),
 \label{eq:Psi-I-relation} \\
 I_s(r) &= \int_0^\infty\!du\int_0^\infty\!dv\,
 u^{s-1}K(u,v)\sin\!\left(r\sqrt{\frac{v}{u}}\right).
 \label{eq:Is-definition}
\end{align}
The convergence strip, differential equation, and confluent-hypergeometric
form of $I_s$ are derived in Appendix~\ref{app:Is-derivation},
subsection~\ref{app:Is-derivation-details}.
Then the regularized zero condition for the function $\zeta_{\rm reg}(s)$,
becomes the global cancellation

\begin{equation}
 \begin{aligned}
 R(s)&= \int_0^\infty dr\,\bigl[e^{-i\pi/8}\Psi_+(r;s) + e^{i\pi/8}\Psi_-(r;s)\bigr]=0.
 \end{aligned}
\label{5.12}
\end{equation}

The two amplitudes have fixed relative phases, and their coherent sum
is the observable on the correlated spectral channel described below.
At the first regularized ordinate, $T\approx14.070052919681542$,
the two phase-weighted integrals are approximately
$-3.39547\times10^{-9}i$ and $+3.39547\times10^{-9}i$, respectively.
The values were checked by quadrature of the Kummer representation
in Eq.~\eqref{eq:app-Is-hypergeometric} at 35 decimal digits. The
two component intensities remain nonzero at their cancellation. This cancellation is reminiscent of
the boundary condition relating the left- and right-moving
fermion components at the boundary in the Dirac--Rindler model of
Ref.~\cite{GS16}. Here, however, Eq.~\eqref{5.12} is a global integral
condition rather than a local boundary condition on the spinor.

\section{Spaces and Hamiltonians}
\label{sec:spaces-Hamiltonians}

In the preceding section, we identified the two components
$\Psi_\pm(r;s)$ of the spinor $\boldsymbol\Psi_s$ from the
Riemann--Siegel integrals. We now construct the physical Hamiltonian
and derive its formal eigenvalue equation for $\boldsymbol\Psi_s$,
keeping $s$ complex within the domain of the integral representation.
The restriction to the critical line follows later from the
self-adjoint realization. The construction uses two auxiliary
spinors. The attenuated spinor
$\boldsymbol\Phi_s$ collects the integral functions $I_s$ and
$I_{1-s}$ before arithmetic dressing, and the reference spinor
$\boldsymbol\Upsilon_s$ consists of the underlying Mellin modes.

The starting point is the differential equation~\eqref{eq:main-Is-ode}
for the integral family $I_s$,
\begin{equation}
 \left[2\pi\partial_r^2+r\partial_r+1-s\right]I_s(r)=0,
 \qquad -1<\Re s<2.
 \label{eq:main-Is-ode}
\end{equation}
Appendix~\ref{app:Is-derivation} derives this equation with explicit
convergence estimates and gives the confluent-hypergeometric expression
of $I_s$. We now organize the construction into three representations:
a reference representation, its attenuated representation obtained by heat
transformation, and its arithmetic dressing.

\subsection{Three state families and Hilbert spaces}
The three generalized spinors are
\begin{equation}
 \begin{aligned}
 \boldsymbol\Upsilon_s(r)&=
 \begin{pmatrix}a_s r^{s-1}\\a_{1-s}r^{-s}\end{pmatrix},\\
 \boldsymbol\Phi_s(r)&=
 \begin{pmatrix}I_s(r)\\I_{1-s}(r)\end{pmatrix},\\
 \boldsymbol\Psi_s(r)&=
 \begin{pmatrix}\widehat c_+(r)I_s(r)\\
 \widehat c_-(r)I_{1-s}(r)\end{pmatrix},
 \end{aligned}
 \label{eq:main-Mellin-spinor}
\end{equation}
where $a_s= (2 \sqrt{\pi})^{-s} \Gamma(1-s/2) $, with removable singularities
understood by continuation. The relation to Mellin modes is initially
established for $0<\Re s<1$ in Appendix~\ref{app:heat-metric}.
The families on $s=1/2+it$ generate generalized spectral representations;
finite-norm states are wave packets, not the individual spinors.
The full spaces have two independent spectral channels. The correlated
integral spinors above select one channel; their two components need
not span the full two-channel space with a common scalar coefficient.

The reference Hilbert space is
\begin{equation}
 \mathcal H_0=L^2((0,\infty),dr)\otimes\mathbb C^2,
 \qquad \langle f,g\rangle_0=\int_0^\infty f^\dagger g\,dr.
 \label{eq:three-reference-space}
\end{equation}
We denote the intermediate attenuated space by $\mathcal H_{\rm att}$.
Attenuation refers to the suppression of high sine-momentum amplitudes
in the reference representation; no equilibrium temperature is assigned.
The attenuated and physical spaces will be the ranges of the transformations
$Q$ and $S=CQ$, equipped with transported inner products. Thus their
Hilbert-space structures include both the underlying sets and their
norms; they are not ranges with the inherited $L^2$ norm.

\subsection{Dilation, attenuation, and arithmetic dressing}
Define the differential operators
\begin{equation}
 A=r\partial_r+\tfrac12,\qquad B=2\pi\partial_r^2.
 \label{eq:fixed-scale-AB}
\end{equation}
For $B$ we use the self-adjoint, nonpositive Dirichlet realization.
The arithmetic multiplication operator and the heat operator are
\begin{equation}
 \begin{aligned}
 C&=\operatorname{diag}(\widehat c_+,\widehat c_-),\\
 Q&=e^{B/2},\qquad S=CQ,\qquad T=S^{-1}=Q^{-1}C^{-1},\\
 \boldsymbol\Phi_s&=Q\boldsymbol\Upsilon_s,\qquad
 \boldsymbol\Psi_s=C\boldsymbol\Phi_s=S\boldsymbol\Upsilon_s.
 \end{aligned}
 \label{eq:main-heat-transform}
\end{equation}
The operator $C$ is fixed by the Riemann--Siegel denominators in
Eq.~\eqref{5.9}; unlike $Q$, it is not defined as a function of $B$.
Both $C$ and $Q$ are bounded and injective in the reference space;
their inverses on their ranges are unbounded. Relations on individual
generalized spinors are interpreted through their spectral extensions.

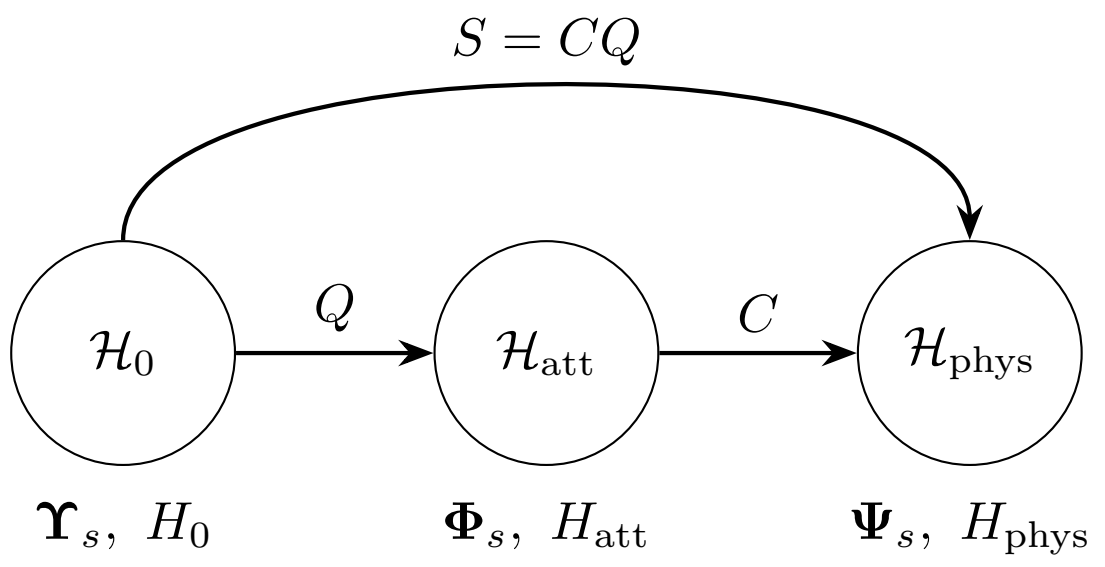
\begin{figure}[t]
 \centering
 \begin{tikzpicture}[>=Stealth,font=\small]
  \node[draw,circle,minimum size=1.48cm,inner sep=1pt] (ref) at (0,0)
   {$\mathcal H_0$};
  \node[draw,circle,minimum size=1.48cm,inner sep=1pt] (att) at (2.8,0)
   {$\mathcal H_{\rm att}$};
  \node[draw,circle,minimum size=1.48cm,inner sep=1pt] (phys) at (5.6,0)
   {$\mathcal H_{\rm phys}$};
  \draw[->,thick] (ref.east) -- node[above] {$Q$} (att.west);
  \draw[->,thick] (att.east) -- node[above] {$C$} (phys.west);
  \draw[->,thick] (ref.north) .. controls +(0,1.35) and +(0,1.35) ..
   node[above] {$S=CQ$} (phys.north);
  \node[below=3pt] at (ref.south) {$\boldsymbol\Upsilon_s,\ H_0$};
  \node[below=3pt] at (att.south) {$\boldsymbol\Phi_s,\ H_{\rm att}$};
  \node[below=3pt] at (phys.south) {$\boldsymbol\Psi_s,\ H_{\rm phys}$};
 \end{tikzpicture}
 \caption{Reference, attenuated, and physical representations. The arrows
 are isometric isomorphisms for the transported inner products.
 Individual critical-line spinors are generalized states; normalizable
 states are wave packets. The upper arrow is the ordered composition
 of heat filtering followed by arithmetic dressing.}
 \label{fig:three-Hilbert-spaces}
\end{figure}

More precisely, set
\begin{equation}
 \begin{aligned}
 \mathcal H_{\rm att}&=\operatorname{Ran}Q,\qquad
 \mathcal H_{\rm phys}=\operatorname{Ran}S,\\
 \langle Qf,Qg\rangle_{\rm att}&=\langle f,g\rangle_0,\\
 \langle Sf,Sg\rangle_{\rm phys}&=\langle f,g\rangle_0.
 \end{aligned}
 \label{eq:three-space-products}
\end{equation}
These are complete Hilbert spaces. The maps
$Q:\mathcal H_0\to\mathcal H_{\rm att}$,
$C:\mathcal H_{\rm att}\to\mathcal H_{\rm phys}$, and
$S:\mathcal H_0\to\mathcal H_{\rm phys}$ are isometric isomorphisms,
and hence unitary between the indicated spaces. They are not unitary
operators in the ordinary reference product. The physical metric is
the closed positive form of $\eta=T^\dagger T$; Appendix~\ref{app:heat-metric}
specifies its domain. Figure~\ref{fig:three-Hilbert-spaces} summarizes
the three representations.

\subsection{Hamiltonians and eigenvalue equations}
The two components of $\boldsymbol\Upsilon_s$ are Mellin modes of
the dilation generator $A=r\partial_r+\tfrac12$. In fact,
$A r^{s-1}=(s-\tfrac12)r^{s-1}$ and
$A r^{-s}=-(s-\tfrac12)r^{-s}$, so
\[
 A\boldsymbol\Upsilon_s=(s-\tfrac12)\sigma^z\boldsymbol\Upsilon_s.
\]
The relation between this first-order equation and the second-order
equation~\eqref{eq:main-Is-ode} follows from the commutator
$[A,B]=-2B$. Since $[B/2,A]=B$ and $[B/2,B]=0$, conjugation by the
heat operator $Q=e^{B/2}$ gives the intertwining identity
$QA=(A+B)Q$ on a suitable core. Applying it to the Mellin modes
through the integral representation of $\boldsymbol\Phi_s
=Q\boldsymbol\Upsilon_s$ gives
\[
 (A+B)\boldsymbol\Phi_s=(s-\tfrac12)\sigma^z\boldsymbol\Phi_s.
\]
The upper component is Eq.~\eqref{eq:main-Is-ode} for $I_s$,
with $B=2\pi\partial_r^2$ supplying the second derivative.
The lower component gives the same equation with $s$ replaced by
$1-s$.

The corresponding equation for the physical spinor follows by applying
$C$ to this last relation. Since $\boldsymbol\Psi_s=C\boldsymbol\Phi_s$
and $C$ commutes with $\sigma^z$, it reads
\[
 C(A+B)C^{-1}\boldsymbol\Psi_s
 =(s-\tfrac12)\sigma^z\boldsymbol\Psi_s.
\]
These formal relations hold for complex $s$ in the common domain of the
representations. Multiplying each by $-i\sigma^z$ gives the eigenvalue
parameter $E(s)=-i(s-\tfrac12)$ and motivates the transported
Hamiltonians
\begin{equation}
 \begin{aligned}
 H_0&=-i\sigma^z A,\\
 H_{\rm att}&=QH_0Q^{-1}=-i\sigma^z(A+B),\\
 H_{\rm phys}&=CH_{\rm att}C^{-1}=SH_0S^{-1}.
 \end{aligned}
 \label{eq:three-Hamiltonians}
\end{equation}
Here and below $H\equiv H_{\rm phys}$ denotes the dressed Hamiltonian.
Their domains are $D(H_0)$, $QD(H_0)$, and $SD(H_0)$, respectively.
The reference domain is explicitly
\begin{equation}
 \begin{aligned}
 D(H_0)&=U^{-1}\!\left[H^1(\mathbb R)\otimes\mathbb C^2\right]\\
 &=\{f\in\mathcal H_0:\ f\in H^1_{\rm loc}(0,\infty)\otimes\mathbb C^2,
 \ Af\in\mathcal H_0\},\\
 (Uf)(x)&=e^{x/2}f(e^x),\qquad x=\log r.
 \end{aligned}
 \label{eq:main-H0-domain}
\end{equation}
Here $H^1(\mathbb R)$ consists of square-integrable functions with a
square-integrable weak first derivative. The closure of $H_0$ initially
defined on $C_c^\infty(0,\infty)\otimes\mathbb C^2$ has this domain.
The endpoints $r=0,\infty$ correspond to $x=-\infty,+\infty$; no
finite-endpoint boundary condition is imposed on $H_0$. In particular,
this domain is different from the Dirichlet domain chosen for the
second-order operator $B$. Domains of the transported Hamiltonians
are defined by $Q$ and $S$, rather than by independently imposing a
new boundary condition on their formal differential expressions.

Their generalized eigenvalue equations are
\begin{equation}
 \begin{aligned}
 H_0\boldsymbol\Upsilon_s&=E(s)\boldsymbol\Upsilon_s,\\
 H_{\rm att}\boldsymbol\Phi_s&=E(s)\boldsymbol\Phi_s,\\
 H_{\rm phys}\boldsymbol\Psi_s&=E(s)\boldsymbol\Psi_s,
 \qquad E(s)=-i(s-\tfrac12).
 \end{aligned}
 \label{eq:spectral}
\end{equation}
In differential form, introducing $\mathbb D=C\partial_rC^{-1}$ gives
\begin{equation}
 \begin{aligned}
 H_{\rm phys}&=-i\sigma^z
 \left(2\pi\mathbb D^2+r\mathbb D+\tfrac12\right),\\
 D_\pm&=\partial_r+
 \frac{\alpha_\pm e^{\alpha_\pm r}}{1+e^{\alpha_\pm r}}.
 \end{aligned}
 \label{eq:fixed-scale-Hamiltonian}
\end{equation}
The first-order term generates opposite dilation flows; the second-order
term is fixed by the Riemann--Siegel kernel. The unitary logarithmic
change $(Uf)(x)=e^{x/2}f(e^x)$ maps $H_0$ to
$-i\sigma^z\partial_x$ on the real line. Its self-adjoint realization
has real absolutely continuous spectrum of multiplicity two. The
transported Hamiltonians are self-adjoint in their respective Hilbert
spaces and have the same spectrum. Spectral membership therefore
requires $s=1/2+it$, giving $E(s)=t$; the formal off-line solutions do
not belong to this self-adjoint spectral representation. The additional
zero condition remains a selection of points in the continuum.

\subsection{Attenuation filtering and the arithmetic filter}
In the sine-transform representation the heat operator multiplies
amplitudes by
\begin{equation}
 (\mathcal F_{\sin}Qf)(k)=e^{-\pi k^2}(\mathcal F_{\sin}f)(k).
 \label{eq:main-heat-filter}
\end{equation}
Thus high spatial frequencies are attenuated more strongly than low
ones. The Gaussian is diagonal in sine momentum, not in the Mellin
spectral parameter $t$. The reference modes $\boldsymbol\Upsilon_s$ are the unfiltered
inputs, and $\boldsymbol\Phi_s=Q\boldsymbol\Upsilon_s$ are the
attenuated generalized modes. No temperature is assigned to either
family. The operator $Q=e^{-\pi K}$, with $K=-B/(2\pi)\geq0$,
is the Dirichlet heat filter: it is an imaginary-time attenuation
operator for the auxiliary Laplacian, rather than a thermal evolution
of the dilation Hamiltonian $H_0$. For finite-norm packets the distinction
between reference attenuation and isometric transport is explicit:
\begin{equation}
 \|Qf\|_0\leq\|f\|_0,\qquad
 \|Qf\|_{\rm att}=\|f\|_0.
 \label{eq:reference-attenuation-vs-transport}
\end{equation}
The name ``attenuated space'' refers to the first comparison; the
transported norm in the second is conserved.

The arithmetic dressing instead multiplies in position space. Its
entries have equal moduli and conjugate phases, with
$\widehat c_\pm(r)\sim2e^{-r/\sqrt2}e^{\pm ir/\sqrt2}$ at large $r$.
Their arithmetic content is exposed by the Mellin transform:
\begin{equation}
 \begin{aligned}
 \mathcal M[\widehat c_\pm](z)
 &=\int_0^\infty r^{z-1}\widehat c_\pm(r)\,dr\\
 &=2\Gamma(z)\alpha_\pm^{-z}(1-2^{1-z})\zeta(z).
 \end{aligned}
 \label{eq:main-C-Mellin}
\end{equation}
For $\Re z>1$ this follows by termwise integration of
$\widehat c_\pm(r)=2\sum_{n\geq1}(-1)^{n-1}e^{-n\alpha_\pm r}$;
it extends to $\Re z>0$, with the singularity at $z=1$ removable.
We use $\log\alpha_\pm=\mp i\pi/4$. The Euler product of $\zeta(z)$
then relates this filter to the primes for $\Re z>1$. This statement
concerns the Mellin transform of $C$; it does not assert an Euler
product for $\zeta_{\rm reg}$.

The arithmetic dressing also has an exact exponential representation,
\begin{equation}
 C=e^{-V_{\rm ar}},\qquad
 V_{\rm ar}=\operatorname{diag}(v_+,v_-),\qquad
 v_\pm=\alpha_\pm r-\log2+\log(1+e^{-\alpha_\pm r}).
 \label{eq:main-arithmetic-generator}
\end{equation}
The logarithm fixes $v_\pm(0)=0$ continuously. To expose its prime
structure, introduce scalar dilations $(\mathsf D_n f)(r)=f(nr)$ and
$h=-i(r\partial_r+1/2)$. Their norm-convergent Euler product gives
\begin{equation}
 \begin{aligned}
 \mathcal Z&=\sum_{n\geq1}\frac{\mathsf D_n}{n}
 =\prod_p(I-\mathsf D_p/p)^{-1}=e^{\mathcal P},\\
 \mathcal P&=\sum_p\sum_{m\geq1}\frac{p^{-3m/2}}m
 e^{im(\log p)h}.
 \end{aligned}
 \label{eq:main-prime-dilations}
\end{equation}
For the exponential seed $f_\alpha(r)=e^{-\alpha r}$ with
$\Re\alpha>0$, the same operator gives
$(\mathcal Z f_\alpha)(r)=-\log(1-e^{-\alpha r})$ for $r>0$.
The generator profile is then
\begin{equation}
 v_\pm(r)=\alpha_\pm r-\log2+
 [(I-\mathsf D_2)\mathcal Z f_{\alpha_\pm}](r).
 \label{eq:main-prime-profile}
\end{equation}
Prime dilations therefore construct the profile of $-\log C$; the
nonlocal operator $\mathcal Z$ and the multiplication filter $C$ are
different objects. In logarithmic coordinates the factors translate by
$m\log p$, recalling primitive-orbit periods and their repetitions in
the periodic-orbit interpretation of the zeta function~\cite{BK99}.
This is an algebraic correspondence, not yet a closed-orbit dynamics
or a trace formula for $H_{\rm phys}$.

The order of the two filters is essential. With the Dirichlet heat
kernel $q(r,u)$ of Appendix~\ref{app:heat-metric},
\begin{equation}
 ([C,Q]f)(r)=\int_0^\infty
 q(r,u)\bigl[C(r)-C(u)\bigr]f(u)\,du.
 \label{eq:main-CQ-commutator}
\end{equation}
Since $C$ is nonconstant, this commutator is nonzero. Consequently
$S=CQ\ne QC$: $S$ is the composite heat-and-arithmetic filter,
whereas $Q$ alone is the heat operator.

\section{PT symmetry}
\label{sec:PT-symmetry}

\subsection{The Riemann--Siegel state family}
In our model the parity is the internal component exchange,
$P=\sigma^x$, and time reversal is complex conjugation, so
$\Theta=\sigma^x\mathcal K$. It is not necessary to reflect the
half-line coordinate. The Hamiltonian and its transported domain obey
\begin{equation}
 \Theta^2=1,\qquad \Theta H_{\rm phys}\Theta^{-1}=H_{\rm phys}.
 \label{eq:PT}
\end{equation}
Indeed, complex conjugation changes the sign of $i$, component exchange
changes the sign of $\sigma^z$, and $A$ and $B$ have real coefficients.
The real heat kernel gives $Q\Theta=\Theta Q$, while
$\widehat c_- =\overline{\widehat c_+}$ gives $C\Theta=\Theta C$.
The reference domain is invariant, as are its images under $Q$ and $S$.
Thus $\Theta$ is antiunitary in all three transported inner products.

For the integral family, conjugation of the defining integrals gives
\begin{equation}
 \Theta\boldsymbol\Phi_s=\boldsymbol\Phi_{1-\bar s},\qquad
 \Theta\boldsymbol\Psi_s=\boldsymbol\Psi_{1-\bar s}.
 \label{eq:PT-integral-family}
\end{equation}
On the critical line $1-\bar s=s$, so the particular integral spinors
are exactly $PT$ invariant. Equivalently,
$a_{1-s}=\overline{a_s}$ there and
$\Theta\boldsymbol\Upsilon_s=\boldsymbol\Upsilon_s$; commuting $\Theta$
through $Q$ and $C$ preserves this property. Their real energies are
therefore consistent with unbroken $PT$. In the present construction,
reality has also been established independently by self-adjoint
transport. Since the full continuum has multiplicity two, real energies
alone would only allow the choice of a $PT$-invariant channel basis;
the explicit conjugation relation proves invariance of the chosen
integral family. Appendix~\ref{app:heat-metric} explains the stronger
complex-linear relation selecting its multiplicity-one channel.

\subsection{Application to the integrated zero condition}
Let
\begin{equation}
 \ell(\boldsymbol\psi)=\int_0^\infty dr\,
 [e^{-i\pi/8}\psi_+(r)+e^{i\pi/8}\psi_-(r)]
 \label{eq:PT-cancellation-functional}
\end{equation}
whenever the integral exists. The opposite coefficient phases give
$\ell(\Theta\boldsymbol\psi)=\overline{\ell(\boldsymbol\psi)}$.
Thus Eq.~\eqref{5.12}, the integrated cancellation condition, is
$PT$ invariant. On the critical line its integral-family solutions
are $PT$-invariant generalized eigenstates with real energies.

For general $s=\sigma+it$ the same analytic family instead obeys
\begin{equation}
 \begin{aligned}
 s'&=1-\bar s=(1-\sigma)+it,\\
 E(s')&=E(s)^*.
 \end{aligned}
 \label{eq:PT-offline-pairs}
\end{equation}
Hence any off-critical-line zeros of Eq.~\eqref{5.12} occur in
reflection-related pairs, corresponding to conjugate formal energies.
The real-spectrum realization excludes these as physical spectral
states, but showing that every analytic zero belongs to that realization
is still required to prove RRH. The cancellation condition by itself
neither establishes spectral reality nor converts the absolutely
continuous spectrum into a discrete one. 

\begin{figure}[h!]
 \centering
 \begin{tikzpicture}[x=1cm,y=.75cm,>=Stealth,font=\small]
  \draw[->] (-1.6,0)--(1.7,0) node[right] {$\Re s$};
  \draw[->,densely dashed] (0,-1.7)--(0,1.1) node[above] {$\Re s=\tfrac12$};
  \fill[orange!80!black] (.65,-.95) circle (1.7pt) node[below right] {$s$};
  \fill[orange!80!black] (-.65,-.95) circle (1.7pt) node[below left] {$1-\bar s$};
  \draw[orange!80!black,<->] (-.55,-.95)--(.55,-.95);
  \begin{scope}[xshift=6cm]
   \draw[->] (-1.6,0)--(1.7,0) node[right] {$\Re E$};
   \draw[->] (0,-1.6)--(0,1.1) node[above] {$\Im E$};
   \fill[orange!80!black] (-.85,-.68) circle (1.7pt) node[left] {$E(s)$};
   \fill[orange!80!black] (-.85,.68) circle (1.7pt) node[left] {$\overline{E(s)}$};
   \draw[orange!80!black,<->] (-.85,-.55)--(-.85,.55);
  \end{scope}
 \end{tikzpicture}
 \caption{Crossed reflection of a hypothetical off-line analytic zero
 and conjugation of its formal energy $E(s)=-i(s-\tfrac12)$.
 The critical line maps to the real spectral axis. The off-axis
 points illustrate the analytic symmetry and carry no physical
 spectral projection.}
 \label{fig:PT-zero-pairs}
\end{figure}
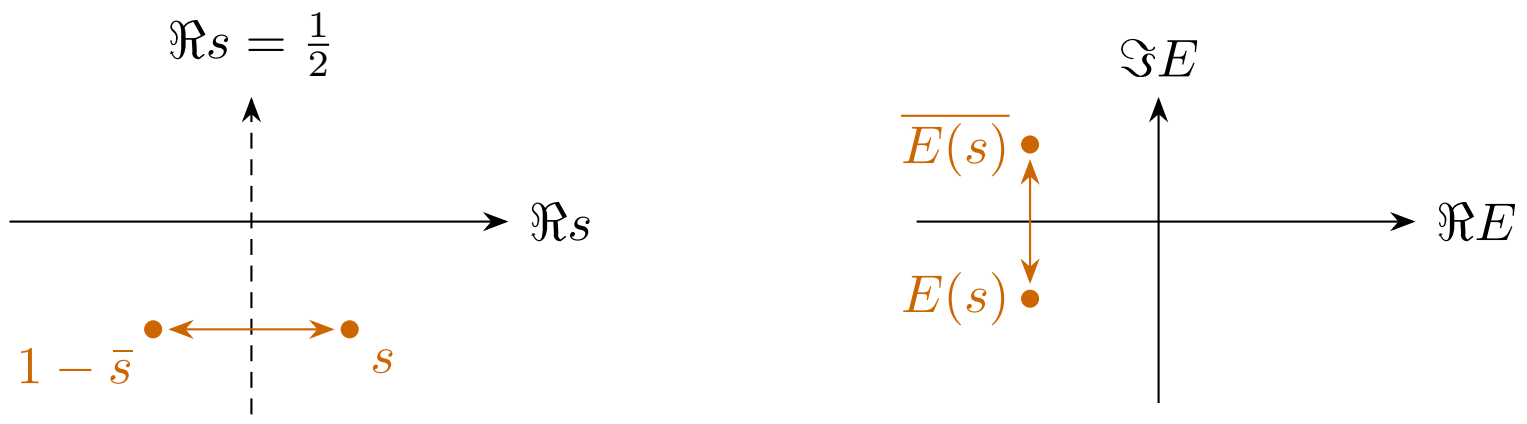

Figure \ref{fig:PT-zero-pairs} summarizes the reflection of the spectral parameter and the conjugation of formal energies. 

\section{Relation to other approaches}
The term $-iA$ is the symmetrically ordered Berry--Keating operator
\cite{BK99}. Here the integral representation supplies its second-order
extension and the global zero-selection amplitude. Unlike the modified
$xp$ models of references
\cite{SierraRodriguezLaguna11,BerryKeatingCompact11}, the second-order correction here is
fixed by the integral kernel, rather than introduced to close classical
orbits. Connes's absorption picture interprets the zeros as missing lines in a continuous spectrum \cite{Connes99}. However, a quantum realization of his version of the $xp$ model attributes the corresponding contribution to a finite-size correction, challenging this interpretation \cite{SierraTownsend08}. In the present construction, Eq.~\eqref{5.12} expresses an integrated amplitude cancellation whose connection to a trace formula remains to be established. The two components
recall Srednicki's two-valued degree of freedom \cite{Srednicki11}, but
internal $PT$ symmetry alone does not establish class-C symmetry.

Bender, Brody, and M\"uller~\cite{BenderBrodyMuller17} connect a
non-Hermitian extension of dilation dynamics to the zeta function through
a boundary condition. Their $PT$-symmetric operator is $iH_{\rm BBM}$:
RH would correspond to maximally broken $PT$ symmetry, with purely
imaginary conjugate-paired eigenvalues of $iH_{\rm BBM}$ and hence real
energies of $H_{\rm BBM}$. In our construction, critical-line states
instead belong to the unbroken-$PT$ sector of $H$.
Yakaboylu's constructions~\cite{Yakaboylu24,Yakaboylu26}
also combine dilation dynamics with nonunitary transformations.
The 2024 proposal relates zeta zeros to a boundary condition,
whereas the 2026 work formulates RH through positivity of an
intertwining operator connecting zero-related spectral sectors.
Our approach instead starts from the exact Riemann--Siegel
integrals and constructs a positive-metric realization with
continuous real spectrum. The regularized zeros are then selected
by integrated amplitude cancellation. The remaining challenge is
to establish that every analytic zero belongs to this physical
spectral realization; positivity of the transported metric alone
does not establish RRH.

\section{Conclusions}
Riemann--Siegel representation defines a regularized zeta function
with reflection symmetry across the critical line. We prove that its
upper critical half-strip is zero-free. In the numerical sample, the
lower critical-line zeros lie close to the ordinary Riemann zeros and
show similar spacing and pair-correlation statistics, broadly
consistent with GUE benchmarks.

The Dirichlet heat transformation and arithmetic dressing give a
self-adjoint Hamiltonian in a positive physical metric. Its spectrum
is real and continuous, and the critical-line integral states are
$PT$ invariant. The regularized critical-line zeros occur where two
integrated amplitudes cancel. These zeros are nodes of a coherent
response within the continuum, reminiscent of Connes's missing
spectral lines~\cite{Connes99}. A discrete Hilbert--P\'olya
realization remains open.

The arithmetic dressing is generated from an exponential seed by a
norm-convergent Euler product of dilations. Prime powers contribute
logarithmic shifts $m\log p$, which resemble repetitions in a
periodic-orbit expansion. A closed-orbit realization and a trace
formula remain open, as do RRH and its conjectured equivalence with RH.
\begin{acknowledgments}

I thank Ignacio Cirac, Jon Keating, Jos\'e Ignacio Latorre, Andr\'{e} LeClair, Jon Merino,
Giuseppe Mussardo, Klaus Richter, Javier Rodríguez-Laguna, Agust\'{\i}n Sabio, Silvia Santalla, 
 Mark Srednicki and Enderalp Yakaboylu for helpful discussions.
I acknowledge financial support from the Spanish grants  PID2024-161474NB-I00,
QUITEMAD-CM TEC2024/COM-84, and  CEX2025-001574-S.

During the preparation of this work, the author used ChatGPT
(OpenAI, GPT-6.2 Sol) to assist with manuscript drafting and
editing, the discussion of mathematical derivations, and the critical
examination of intermediate arguments. The author directed this use,
reviewed and revised the resulting material, and takes full
responsibility for the scientific content and conclusions.  

\end{acknowledgments}

\appendix

\section{Distributional Limit of the Riemann--Siegel Contours}
\label{app:distributional-limit}

We derive the regular and singular contributions to the completed zeta
function, keeping $0<\Re s<1$ throughout the distributional calculation.
The starting point is Eq.~\eqref{3.6}, written as
\begin{equation}
\Lambda(s)=
\pi^{-\frac{s}{2}}\Gamma\!\left(\frac{s}{2}\right)F_+(s,\epsilon)
+\pi^{\frac{s-1}{2}}\Gamma\!\left(\frac{1-s}{2}\right)
F_-(1-s,\epsilon),
\label{4.1}
\end{equation}
where the two contour integrals take the form
\begin{align}
F_+(s,\epsilon)
&=\frac{\alpha_+}{2i}\int_{-\infty}^{\infty}dx\,
z_+^{-s}\,
\frac{e^{-i\pi z_+^2}}{\sin(\pi z_+)},
\label{4.2}\\
F_-(1-s,\epsilon)
&=-\frac{\alpha_-}{2i}\int_{-\infty}^{\infty}dx\,
z_-^{s-1}\,
\frac{e^{i\pi z_-^2}}{\sin(\pi z_-)},
\label{4.3}
\end{align}

Here $\alpha_\pm=e^{\mp i\pi/4}$. The displaced contours are parametrized by
\begin{align}
 z_+(x,\epsilon)&=\alpha_+(x+i\epsilon),\quad x:-\infty\to\infty,
 \label{3.7}\\
 z_-(x,\epsilon)&=\alpha_-(x-i\epsilon),\quad x:\infty\to-\infty.
 \label{3.8}
\end{align}
with $0<\epsilon<1/\sqrt{2}$. The reversed orientation of the second contour
accounts for the minus sign in Eq.~\eqref{4.3}. Each contour crosses the
real axis at $\sqrt{2} \epsilon$, between zero and one. Contour deformation
makes the integrals independent of $\epsilon$ within this range, whereas
the limit $\epsilon\to0^+$ requires a distributional prescription.
We use the principal logarithm,
\begin{equation}
z^{-s}=e^{-s\log z},\quad
z^{s-1}=e^{(s-1)\log z},\quad -\pi<\arg z<\pi.
\label{4.4}
\end{equation}

\subsection{Pole subtraction and regular contributions}

On the limiting contours, the only zero of $\sin(\pi z_\pm)$ is at the
origin. We isolate its contribution using
\begin{equation}
\csc(\pi z)=\frac{1}{\pi z}+h(z),
\qquad h(z)\ \text{analytic at }z=0,
\label{4.5}
\end{equation}

The corresponding singular factors behave as
\begin{equation}
\frac{z^{-s}}{\sin(\pi z)}
\sim\frac{z^{-s-1}}{\pi},
\qquad
\frac{z^{s-1}}{\sin(\pi z)}
\sim\frac{z^{s-2}}{\pi}.
\label{4.6}
\end{equation}

Both are nonintegrable at the origin in the critical strip. Their boundary
values are defined by analytic continuation, equivalently by the Hadamard
finite part. This prescription gives
\begin{equation}
F_\pm=F_{\pm,\mathrm{reg}}+F_{\pm,\mathrm{sing}}.
\label{4.7}
\end{equation}

The regular terms contain the odd function
$h(z)=\csc(\pi z)-1/(\pi z)$, which is $O(z)$ at the origin.
They are therefore locally integrable. To reduce them to the positive
half-line, we use
\begin{align}
(x+i0)^{-s}
&=\Theta(x)x^{-s}
+\Theta(-x)|x|^{-s}e^{-i\pi s},
\label{4.10}\\
(x-i0)^{s-1}
&=\Theta(x)x^{s-1}
+\Theta(-x)|x|^{s-1}e^{-i\pi(s-1)}.
\label{4.11}
\end{align}

For the plus contour, splitting the integral at $x=0$ and using the
oddness of $h$ gives the phase factor
$\alpha_+^{1-s}(1-e^{-i\pi s})/(2i)
=e^{-i\pi(s+1)/4}\sin(\pi s/2)$.
The corresponding factor for the minus contour is
$-\alpha_-^s(1-e^{-i\pi(s-1)})/(2i)
=e^{-i\pi(s-2)/4}\sin(\pi(1-s)/2)$.
Thus
\begin{align}
F_{+,\mathrm{reg}}(s)
={}&
e^{-\frac{i\pi}{4}(s+1)}
\sin\!\left(\frac{\pi s}{2}\right)
\notag\\
\times &
\int_0^\infty dx\,x^{-s}e^{-\pi x^2}
\left[
\csc(\pi\alpha_+x)
-\frac{1}{\pi\alpha_+x}
\right],
\label{4.8}\\[1mm]
F_{-,\mathrm{reg}}(1-s)
={}&
e^{-\frac{i\pi}{4}(s-2)}
\sin\!\left(\frac{\pi(1-s)}{2}\right)
\notag\\
\times &
\int_0^\infty dx\,x^{s-1}e^{-\pi x^2}
\left[
\csc(\pi\alpha_-x)
-\frac{1}{\pi\alpha_-x}
\right].
\label{4.9}
\end{align}

The regular part of the completed function is consequently
\begin{align}
\Lambda_{\mathrm{reg}}(s)
={}&
\pi^{-s/2}\Gamma\!\left(\frac{s}{2}\right)
F_{+,\mathrm{reg}}(s)
\notag\\
+&
\pi^{(s-1)/2}
\Gamma\!\left(\frac{1-s}{2}\right)
F_{-,\mathrm{reg}}(1-s),
\label{4.20}
\end{align}

The Gaussian controls the integrals at infinity. At the origin their
integrands are respectively $O(x^{1-\Re s})$ and $O(x^{\Re s})$, so the
regular integral representations extend to $-1<\Re s<2$.

\subsection{Singular contributions}

The subtracted pole yields the distributional pairings
\begin{align}
F_{+,\mathrm{sing}}(s)
&=
\frac{\alpha_+^{-s}}{2\pi i}
\left\langle (x+i0)^{-s-1},e^{-\pi x^2}\right\rangle,
\label{4.12}\\
F_{-,\mathrm{sing}}(1-s)
&=
-\frac{\alpha_-^{s-1}}{2\pi i}
\left\langle (x-i0)^{s-2},e^{-\pi x^2}\right\rangle.
\label{4.13}
\end{align}

For a Schwartz test function $\varphi$, distributional differentiation
gives
\begin{align}
\left\langle(x+i0)^{-s-1},\varphi\right\rangle
&=\frac1s\int_{\mathbb R}dx\,
(x+i0)^{-s}\varphi'(x),
\label{4.14}\\
\left\langle(x-i0)^{s-2},\varphi\right\rangle
&=\frac1{1-s}\int_{\mathbb R}dx\,
(x-i0)^{s-1}\varphi'(x).
\label{4.15}
\end{align}

The right-hand sides are ordinary locally integrable expressions for
$0<\Re s<1$. In particular, for $\varphi(x)=e^{-\pi x^2}$, the boundary
values in Eqs.~\eqref{4.10}--\eqref{4.11} give
\begin{align}
&\left\langle(x+i0)^{-s-1},e^{-\pi x^2}\right\rangle
\nonumber\\
&\qquad=-\frac{2\pi}{s}(1-e^{-i\pi s})
\int_0^\infty x^{1-s}e^{-\pi x^2}\,dx,
\label{eq:app-singular-plus-moment}\\[1mm]
&\left\langle(x-i0)^{s-2},e^{-\pi x^2}\right\rangle
\nonumber\\
&\qquad=-\frac{2\pi}{1-s}(1-e^{-i\pi(s-1)})
\int_0^\infty x^s e^{-\pi x^2}\,dx.
\label{eq:app-singular-minus-moment}
\end{align}
Using the Gaussian moments
\begin{align}
\int_0^\infty dx\,x^{1-s}e^{-\pi x^2}
&=\frac{\pi^{s/2}}{2\pi}
\Gamma\!\left(1-\frac{s}{2}\right),
\label{4.16a}\\
\int_0^\infty dx\,x^s e^{-\pi x^2}
&=\frac{\pi^{(1-s)/2}}{2\pi}
\Gamma\!\left(\frac{1+s}{2}\right),
\label{4.16b}
\end{align}
and Euler's reflection formula,
\begin{equation}
\Gamma(z)\Gamma(1-z)=\frac{\pi}{\sin(\pi z)}
\label{4.16}
\end{equation}
we obtain
\begin{align}
F_{+,\mathrm{sing}}(s)
&=
-\frac{e^{-i\pi s/4}}
{s\,\pi^{-s/2}\Gamma(s/2)},
\label{4.17}\\
F_{-,\mathrm{sing}}(1-s)
&=
-\frac{e^{-i\pi(s-1)/4}}
{(1-s)\,\pi^{-(1-s)/2}\Gamma((1-s)/2)}.
\label{4.18}
\end{align}

Combining these expressions with the gamma prefactors in
Eq.~\eqref{4.1} gives
\begin{equation}
\Lambda_{\mathrm{sing}}(s)
=-\frac{e^{-i\pi s/4}}{s}
 -\frac{e^{-i\pi(s-1)/4}}{1-s},
\label{eq:app-Lambda-singular}
\end{equation}
in agreement with Eq.~\eqref{4.21}. Together with
Eq.~\eqref{4.20}, this yields
$\Lambda=\Lambda_{\mathrm{reg}}+\Lambda_{\mathrm{sing}}$.
The prescription fixes the finite-part convention and the phases of both
contributions.

\section{Properties and zeros of the regularized zeta function}
\label{app:regularized-properties}

For completeness, we recall the definition
\begin{equation}
\zeta_{\mathrm{reg}}(s)
:=
\frac{\pi^{s/2}}{\Gamma(s/2)}
\Lambda_{\mathrm{reg}}(s),
\label{4.22}
\end{equation}
and, similarly,
\begin{align}
\Delta(s)
&:=
\frac{\pi^{s/2}}{\Gamma(s/2)}
\Lambda_{\mathrm{sing}}(s) \label{4.23} \\
&=
\frac{\pi^{s/2}}{\Gamma(s/2)}
\left[
-\frac{e^{-i\pi s/4}}{s}
-\frac{e^{-i\pi(s-1)/4}}{1-s}
\right].
\nonumber
\end{align}

These definitions yield the exact additive decomposition
\begin{equation}
\zeta(s)
=
\zeta_{\mathrm{reg}}(s)
+
\Delta(s).
\label{eq:appendix-zeta-decomposition}
\end{equation}

The functional equation of the Riemann zeta function can then be written
entirely in terms of the two contributions as
\begin{equation}
\begin{split}
&\pi^{-s/2}\Gamma\!\left(\frac{s}{2}\right)
 \bigl[\zeta_{\mathrm{reg}}(s)+\Delta(s)\bigr]
\\
&\quad=
\pi^{-(1-s)/2}\Gamma\!\left(\frac{1-s}{2}\right)
 \bigl[\zeta_{\mathrm{reg}}(1-s)+\Delta(1-s)\bigr].
\end{split}
\label{eq:zeta-reg-Delta-functional}
\end{equation}
Likewise, Schwarz reflection becomes
\begin{equation}
\begin{split}
&\zeta_{\mathrm{reg}}(s)+\Delta(s)
\\
&\qquad=
\zeta_{\mathrm{reg}}^{*}(s^{*})+\Delta^{*}(s^{*}).
\end{split}
\label{eq:zeta-reg-Delta-Schwarz}
\end{equation}
The explicit expression~\eqref{4.23} also satisfies the completed
cross-reflection identity
\begin{equation}
\begin{split}
&\pi^{-s/2}\Gamma\!\left(\frac{s}{2}\right)\Delta(s)
\\
&\quad=
\pi^{-(1-s)/2}\Gamma\!\left(\frac{1-s}{2}\right)
\Delta^{*}(1-s^{*}).
\end{split}
\label{eq:Delta-cross-reflection}
\end{equation}
Combining Eqs.~\eqref{eq:zeta-reg-Delta-functional}--
\eqref{eq:Delta-cross-reflection}  therefore gives
\begin{equation}
\begin{split}
&\pi^{-s/2}\Gamma\!\left(\frac{s}{2}\right)
\zeta_{\mathrm{reg}}(s)
\\
&\quad=
\pi^{-(1-s)/2}\Gamma\!\left(\frac{1-s}{2}\right)
\zeta_{\mathrm{reg}}^{*}(1-s^{*}).
\end{split}
\label{eq:zeta-reg-cross-reflection}
\end{equation}
Inside the critical strip the gamma prefactors are finite and nonzero.
Consequently, if $\zeta_{\mathrm{reg}}(s_0)=0$, then
\begin{equation}
 \zeta_{\mathrm{reg}}(1-s_0^{*})=0.
 \label{eq:zeta-reg-zero-pairing}
\end{equation}

The zeros of $\zeta_{\mathrm{reg}}$ in the critical strip therefore occur
in reflection pairs,
\begin{equation}
\left\{s_0,\,1-s_0^{*}\right\}.
\end{equation}
If $\operatorname{Re}s_0=\tfrac12$, the zero is invariant under this
reflection, and the pair collapses to a single point. This is in contrast
to the zeros of $\zeta(s)$, which, owing to both the functional equation
and Schwarz reflection, may occur in quartets,
\begin{equation}
\left\{s_0,\,s_0^{*},\,1-s_0,\,1-s_0^{*}\right\},
\end{equation}
when $\operatorname{Re}s_0\neq\tfrac12$.

At the boundary and outside the critical strip, the zeros inherited from
the reciprocal gamma factor require separate treatment.  Directly, 
Eq.  \eqref{4.23}, yields $\Delta(0)= - 1/2$  and since $\zeta(0) = - 1/2$, Eq. \eqref{eq:appendix-zeta-decomposition}, implies
\begin{equation}
 \zeta_{\mathrm{reg}}(0)=0,
 \qquad
 \zeta_{\mathrm{reg}}(-2n)=0,
 \quad n=1,2,\ldots .
 \label{eq:zeta-reg-trivial-zeros}
\end{equation}
The upper critical strip contains no zeros by the analytic
bound~\eqref{eq:upper-strip-bound}, proved in
Appendix~\ref{app:upper-strip-zero-free}. In contrast, lower-half-strip
zeros occur, as illustrated by the critical-line zeros in
Table~\ref{tab:zero-ordinates}. Thus the one-sided distribution is
established analytically for this regularization, whereas the tabulated
locations and their comparison with the Riemann ordinates are numerical.

For $s=\sigma+it$, with $0<\sigma<1$, Stirling's formula gives
\begin{equation}
\begin{split}
&\left|\zeta(\sigma+it)
-\zeta_{\mathrm{reg}}(\sigma+it)\right|
\\[2mm]
&\quad=
\begin{cases}
O_\sigma\!\left(t^{-(1+\sigma)/2}e^{\pi t/2}\right),
& t\to+\infty,
\\[2mm]
O_\sigma\!\left(|t|^{-(1+\sigma)/2}\right),
& t\to-\infty.
\end{cases}
\end{split}
\label{4.24}
\end{equation}
Here \(O_\sigma\) means that the implicit constants may depend on the
fixed value of \(\sigma\), but are independent of \(t\). In particular,
on the critical line,
\begin{equation}
\left|
\zeta\!\left(\frac12+it\right) - \zeta_{\mathrm{reg}}\!\left(\frac12+it\right)\right|
=
O\!\left(|t|^{-3/4}\right),
\qquad t\to-\infty.
\label{4.25}
\end{equation}
The function approximation is one-sided. Independently, Schwarz reflection
relates positive and negative Riemann zeros; this does not reconstruct
the exact Riemann zeros from regularized zeros.  The Schwarz reflection property,
in the form of Eq.~\eqref{eq:zeta-reg-Delta-Schwarz}, implies that every
Riemann zero \(\frac12-i\gamma_n\), with \(\gamma_n>0\), is the complex conjugate
of the zero \(\frac12+i\gamma_n\). Hence the negative critical-line zeros determine
the corresponding positive Riemann ordinates. Introducing the positive
spectral parameter
\begin{equation}
\mathcal E=-t>0,
\end{equation}
the limit \(t\to-\infty\) becomes the high-energy limit \(\mathcal E\to+\infty\).
The numerical comparison and the definitions of the positive ordinate
magnitudes are given in Table~\ref{tab:zero-ordinates} in the main text.
The function estimate~\eqref{4.25} alone does not establish a one-to-one
correspondence between all regularized and Riemann zeros.

\section{Zero-free upper critical strip}
\label{app:upper-strip-zero-free}

We prove the uniform bound~\eqref{eq:upper-strip-bound} and its
zero-free consequence~\eqref{eq:upper-strip-zero-free}.

Define
\begin{equation}
 \psi(x)=\sum_{n=1}^{\infty}e^{-\pi n^2x},\qquad x>0.
\end{equation}
The classical theta-function representation gives~\cite[Sec.~2.6, p.~22]{Titchmarsh86}
\begin{equation}
 \Lambda(s)=-\frac1s-\frac1{1-s}+J(s),
 \label{eq:theta-representation}
\end{equation}
where
\begin{equation}
 J(s)=\int_1^\infty\psi(x)
 \left[x^{s/2-1}+x^{(1-s)/2-1}\right]dx.
 \label{eq:J-definition}
\end{equation}
The integral defining $J$ converges locally uniformly for all complex
$s$ and hence defines an entire function.
For completeness, Eq.~\eqref{eq:theta-representation} follows by
splitting the Mellin integral
\begin{equation}
 \Lambda(s)=\int_0^\infty\psi(x)x^{s/2-1}\,dx,
 \qquad \Re s>1,
\end{equation}
at $x=1$, and applying the theta transformation
\begin{equation}
 \psi(x)=x^{-1/2}\psi(1/x)+\frac{x^{-1/2}-1}{2}
\end{equation}
to the integral over $(0,1)$. The elementary terms give
$1/(s-1)-1/s$, and the remaining integral becomes the second term in
Eq.~\eqref{eq:J-definition}. Analytic continuation then yields
Eq.~\eqref{eq:theta-representation} away from $s=0,1$.

Set $a=\pi/4$. Combining the regularization in Eqs.~\eqref{4.19} and \eqref{4.21} with
Eq.~\eqref{eq:theta-representation}, we obtain
\begin{equation}
 \Lambda_{\mathrm{reg}}(s)
 =J(s)+\frac{e^{-ias}-1}{s}
       +\frac{e^{-ia(s-1)}-1}{1-s}.
\end{equation}
The two explicit terms have the integral representations
\begin{align}
 \frac{e^{-ias}-1}{s}&=-i\int_0^a e^{-ivs}\,dv,\\
 \frac{e^{-ia(s-1)}-1}{1-s}&=i\int_0^a e^{-iv(s-1)}\,dv.
\end{align}
Their sum therefore gives the exact identity
\begin{equation}
 \Lambda_{\mathrm{reg}}(s)
 =J(s)-2\int_0^{\pi/4}\sin(v/2)e^{-iv(s-1/2)}\,dv.
 \label{eq:finite-integral}
\end{equation}

Let $s=\sigma+it$, with $0<\sigma<1$ and $t\geq0$. Taking real parts,
\begin{equation}
 \begin{aligned}
 \Re\Lambda_{\mathrm{reg}}(\sigma+it)
 &=\Re J(\sigma+it)\\
 &\quad-2\int_0^{\pi/4}\sin(v/2)e^{vt}
 \cos\!\left(v(\sigma-\tfrac12)\right)dv.
 \end{aligned}
 \label{eq:real-part}
\end{equation}
For $x\geq1$, both $x^{\sigma/2-1}$ and
$x^{(1-\sigma)/2-1}$ are at most one. Since $\psi(x)>0$,
\begin{align}
 \Re J(s)\leq |J(s)|
 &\leq2\int_1^\infty\psi(x)\,dx\\
 &=\frac2\pi\sum_{n=1}^{\infty}
       \frac{e^{-\pi n^2}}{n^2}\\
 &\leq\frac2\pi\sum_{n=1}^{\infty}e^{-\pi n}
 =\frac{2}{\pi(e^\pi-1)}.
 \label{eq:J-bound}
\end{align}
The exchange of the positive sum and integral is justified by Tonelli's
theorem; the last inequality uses $n^2\geq n$ and $n^{-2}\leq1$.

On the other hand, $t\geq0$ implies $e^{vt}\geq1$. Moreover,
$|\sigma-1/2|\leq1/2$ implies
\begin{equation}
 \cos\!\left(v(\sigma-\tfrac12)\right)\geq\cos(v/2)>0,
 \qquad 0\leq v\leq\pi/4.
\end{equation}
It follows that
\begin{align}
 &2\int_0^{\pi/4}\sin(v/2)e^{vt}
       \cos\!\left(v(\sigma-\tfrac12)\right)dv\\
 &\qquad\geq2\int_0^{\pi/4}\sin(v/2)\cos(v/2)\,dv\\
 &\qquad=\int_0^{\pi/4}\sin v\,dv
 =1-\frac1{\sqrt2}.
 \label{eq:finite-integral-bound}
\end{align}
Combining Eqs.~\eqref{eq:real-part}, \eqref{eq:J-bound}, and
\eqref{eq:finite-integral-bound} proves Eq.~\eqref{eq:upper-strip-bound}.
The upper bound is approximately $-0.26413984$, so
$\Lambda_{\mathrm{reg}}(s)$ cannot vanish in the stated region.
Finally, $\pi^{s/2}/\Gamma(s/2)$ is finite and nonzero in the open
critical strip. Equation~\eqref{eq:zeta-reg-Delta-decomposition} therefore establishes the
same zero-free region for $\zeta_{\mathrm{reg}}$.

The restriction $t\geq0$ enters through $e^{vt}\geq1$. For $t<0$,
this lower bound fails, and the proof does not exclude lower-half-strip
zeros or establish RRH.


\section{Local perturbative correspondence between the two zero sets}
\label{app:perturbative-zero-correspondence}

The exact decomposition $\Lambda=\Lambda_{\rm reg}+\Lambda_{\rm sing}$
allows a local comparison of the zeros. We distinguish a linearized
root estimate from existence of a true nearby zero, and from a global
equivalence between RH and RRH. All neighborhoods below avoid the poles
at $s=0,1$, and primes denote differentiation with respect to $s$.

\subsection{Reflection symmetry and reality on the critical line}

Let $\mathcal J s=1-\bar s$. The completed Riemann function satisfies
$\Lambda(\mathcal J s)=\overline{\Lambda(s)}$. Directly from
Eq.~\eqref{4.21}, the singular term obeys the same identity, and hence
so does $\Lambda_{\rm reg}$. For any of these three functions $f$,
\begin{equation}
 f(\tfrac12-iT)\in\mathbb R,\qquad
 f'(\tfrac12-iT)\in i\mathbb R.
 \label{eq:app-line-reality}
\end{equation}
The derivative statement follows by differentiating the real function
$f(\tfrac12-iT)$ with respect to $T$. More explicitly,
\begin{equation}
 \Lambda_{\rm sing}(\tfrac12-iT)
 =-e^{-\pi T/4}
 \frac{\cos(\pi/8)+2T\sin(\pi/8)}{T^2+\tfrac14}.
 \label{eq:app-singular-on-line}
\end{equation}
For $T>0$ this is strictly negative; in particular, the two completed
functions do not have a common zero on the lower critical line.

\subsection{Linearized displacements in both directions}

Suppose $\Lambda_{\rm reg}(s_0)=0$, with
$s_0=\tfrac12-iT_0$ and $T_0>0$. A candidate Riemann zero
$s_1=s_0+h$ satisfies
\begin{equation}
 0=\Lambda_{\rm sing}(s_0)+h\Lambda'(s_0)
   +O(h^2).
 \label{eq:app-forward-taylor}
\end{equation}
If $\Lambda'(s_0)\ne0$, its first-order estimate is
\begin{equation}
 h_{\rm lin}=-\frac{\Lambda_{\rm sing}(s_0)}{\Lambda'(s_0)}
 \in i\mathbb R.
 \label{eq:app-forward-shift}
\end{equation}
Thus the predicted displacement is tangent to the critical line.
If a true zero exists with displacement $h$ inside a disk on which
$|\Lambda''|\le M$, Taylor's formula gives the explicit error bound
\begin{equation}
 |h-h_{\rm lin}|
 \le \frac{M}{2|\Lambda'(s_0)|}|h|^2.
 \label{eq:app-newton-remainder}
\end{equation}
This estimate is conditional on existence and does not itself establish
a zero near the linear prediction.

Conversely, let $\Lambda(s_1)=0$ on the lower critical line and seek a
regularized zero $s_0=s_1+k$. Since
$\Lambda_{\rm reg}(s_1)=-\Lambda_{\rm sing}(s_1)$,
\begin{equation}
 0=-\Lambda_{\rm sing}(s_1)
    +k\Lambda_{\rm reg}'(s_1)+O(k^2).
\end{equation}
For $\Lambda_{\rm reg}'(s_1)\ne0$ this yields
\begin{equation}
 k_{\rm lin}=\frac{\Lambda_{\rm sing}(s_1)}
                   {\Lambda_{\rm reg}'(s_1)}\in i\mathbb R.
 \label{eq:app-reverse-shift}
\end{equation}
With the common convention $h=s_1-s_0=-k$, the reverse estimate is
$h\simeq-\Lambda_{\rm sing}(s_1)/\Lambda_{\rm reg}'(s_1)$.
These estimates are evaluated at different endpoints and are not an
exact identity between their Newton steps. In both directions they
predict critical-line locations, provided the relevant zero exists
and the linear approximation is accurate.

\vspace{0.2cm}

\section{Integral states and their differential equation}
\label{app:Is-derivation}

\subsection{Integral representation and hypergeometric form of $I_s$}
\label{app:Is-derivation-details}

\subsubsection{Reduction to a Gaussian sine transform}
Start from Eq.~\eqref{eq:Is-definition}, and return to $u=p/x$ and
$v=px$, with $p,x>0$. Since $du\,dv=2u\,dp\,dx$, we obtain
\begin{equation}
 \begin{aligned}
 I_s(r)&=2\int_0^\infty dp\,p^s e^{-\pi p^2}
 \int_0^\infty dx\,x^{-s}e^{-\pi x^2}\sin(rx)\\
 &=b_s\int_0^\infty dx\,x^{-s}e^{-\pi x^2}\sin(rx),\\
 b_s&=\pi^{-(s+1)/2}\Gamma\!\left(\frac{s+1}{2}\right).
 \end{aligned}
 \label{eq:app-Is-factorization}
\end{equation}
The $p$ integral requires $\Re s>-1$. At $x=0$, the sine makes the
integrand $O(x^{1-\Re s})$, requiring $\Re s<2$; Gaussian decay
controls both infinities. Thus the double integral converges absolutely
for $-1<\Re s<2$ at finite $r$, and defines a holomorphic function of
$s$ there. Differentiation with respect to $r$ is justified by dominated
convergence on compact $r$ intervals. The same strip accommodates
$I_{1-s}$.

\subsubsection{Detailed integration-by-parts derivation}
For $r$ real and $-1<\Re s<2$, the boundary terms in
\begin{equation}
 \int_0^\infty \partial_x
 \left[x^{1-s}e^{-\pi x^2}\sin(rx)\right]dx=0
 \label{eq:app-Is-total-derivative}
\end{equation}
vanish: the bracket is $O(x^{2-\Re s})$ at zero and decays
Gaussianly at infinity. Expanding the derivative and multiplying by
$b_s$ gives
\begin{equation}
 \begin{aligned}
 0={}&(1-s)I_s(r)
 -2\pi b_s\int_0^\infty x^{2-s}e^{-\pi x^2}\sin(rx)\,dx\\
 &+r b_s\int_0^\infty x^{1-s}e^{-\pi x^2}\cos(rx)\,dx\\
 ={}&(1-s)I_s(r)+2\pi I_s''(r)+rI_s'(r).
 \end{aligned}
 \label{eq:app-Is-ode-derived}
\end{equation}
This proves Eq.~\eqref{eq:main-Is-ode} without assuming the critical
line. Equivalently, the double-integral kernel satisfies
$(u\partial_u-v\partial_v)K=2\pi(v/u)K$; integrating this identity
against the Mellin-sine factor gives the same equation.

\subsubsection{Confluent hypergeometric solution and normalization}
Expanding $\sin(rx)$ and using
\begin{equation}
 \int_0^\infty x^{2n+1-s}e^{-\pi x^2}dx
 =\tfrac12\pi^{-n-1+s/2}\Gamma(n+1-s/2)
\end{equation}
yields
\begin{equation}
 I_s(r)=d_s r\,
 {}_1F_1\!\left(1-\frac s2;\frac32;-\frac{r^2}{4\pi}\right),
 \label{eq:app-Is-hypergeometric}
\end{equation}
where
\begin{equation}
 d_s=\frac{\Gamma((s+1)/2)\Gamma(1-s/2)}{2\pi^{3/2}},
 \qquad I_s(0)=0,\quad I_s'(0)=d_s.
 \label{eq:app-Is-initial-data}
\end{equation}
Termwise integration is justified for finite complex $r$ by the bound
$e^{-\pi x^2+|r|x}$ at infinity and the same integrability condition
at zero. The identity
$(2n+1)!=4^n n!(3/2)_n$ converts the series to ${}_1F_1$.

The differential equation gives a second verification. Writing
$w=-r^2/(4\pi)$ and $I_s(r)=rF(w)$ reduces it to Kummer's equation
\begin{equation}
 wF''+(\tfrac32-w)F'-(1-s/2)F=0.
 \label{eq:app-Is-Kummer}
\end{equation}
A convenient local basis of solutions of the original equation is
\begin{equation}
 \begin{aligned}
 y_{\rm even}(r)&={}_1F_1\!\left(\frac{1-s}{2};\frac12;
 -\frac{r^2}{4\pi}\right),\\
 y_{\rm odd}(r)&=r\,{}_1F_1\!\left(1-\frac s2;\frac32;
 -\frac{r^2}{4\pi}\right).
 \end{aligned}
\end{equation}
The sine integral selects the odd solution, and its derivative at the
origin fixes the coefficient $d_s$. Kummer's transformation also gives
\begin{equation}
 I_s(r)=d_s r e^{-r^2/(4\pi)}
 {}_1F_1\!\left(\frac{s+1}{2};\frac32;\frac{r^2}{4\pi}\right).
 \label{eq:app-Is-hypergeometric-alternative}
\end{equation}
For the other component, replace $s$ by $1-s$ throughout. These formulas
hold in the convergence strip of the original integral and supply its
meromorphic continuation in $s$ outside that strip.

\section{Dirichlet heat transformation, physical metric, and generalized states}
\label{app:heat-metric}

This appendix constructs the physical Hilbert space by transporting the
self-adjoint dilation generator with the Dirichlet heat operator and the
dressing matrix. It distinguishes normalizable wave packets from
individual generalized eigenfunctions and separates the reality of the
physical continuum from its selection by the zero condition.

\subsection{The Dirichlet heat operator and the intertwining relation}

Let $\mathcal H_0=L^2((0,\infty),dr)\otimes\mathbb C^2$ and let
$B=2\pi\partial_r^2$ act identically on the components. Its scalar
Dirichlet realization has domain
\begin{equation}
 D(B)=H^2(0,\infty)\cap H_0^1(0,\infty).
 \label{eq:app-B-domain}
\end{equation}
It is self-adjoint and nonpositive. The unitary sine transform
\begin{equation}
 (\mathcal F_{\sin}f)(k)=\sqrt{\frac2\pi}
 \int_0^\infty\sin(kr)f(r)\,dr
\end{equation}
gives
\begin{equation}
 \mathcal F_{\sin}B\mathcal F_{\sin}^{-1}=-2\pi k^2,
 \qquad \mathcal F_{\sin}Q\mathcal F_{\sin}^{-1}=e^{-\pi k^2},
 \quad Q=e^{B/2}.
 \label{eq:app-heat-multiplier}
\end{equation}
Consequently $Q$ is bounded, $\|Q\|=1$, and injective: its multiplier is
strictly positive almost everywhere. Since $Q=Q^\dagger$ and
$\ker Q=\{0\}$, its range is dense. It is not surjective; its inverse
has the unbounded multiplier $e^{\pi k^2}$ and domain
\begin{equation}
 D(Q^{-1})=\{g\in L^2:e^{\pi k^2}(\mathcal F_{\sin}g)(k)\in L^2\}
 =\operatorname{Ran}Q.
 \label{eq:app-inverse-heat-domain}
\end{equation}

On a smooth core, $[A,B]=-2B$ yields
\begin{equation}
 QA=(A+B)Q,\qquad A+B=QAQ^{-1}.
 \label{eq:app-heat-intertwining}
\end{equation}
This can also be verified with the Dirichlet heat kernel
\begin{equation}
 q(r,u)=\frac1{2\pi}
 \left[e^{-(r-u)^2/(4\pi)}-e^{-(r+u)^2/(4\pi)}\right].
 \label{eq:app-heat-kernel}
\end{equation}
The second identity in \eqref{eq:app-heat-intertwining} is an identity
on the transported domain, not an assertion that $Q^{-1}$ is bounded
or defined on all of $\mathcal H_0$.

\subsection{The intermediate attenuated Hilbert space}

Set $\mathcal H_{\rm att}=\operatorname{Ran}Q$ and define
\begin{equation}
 \langle\phi,\chi\rangle_{\rm att}
 =\langle Q^{-1}\phi,Q^{-1}\chi\rangle_0.
 \label{eq:app-heat-product}
\end{equation}
Since $Q$ is injective, this is unambiguous, and $Q$ is an isometric
isomorphism onto this space. Completeness follows by transporting
Cauchy sequences back to $\mathcal H_0$. Thus
$H_{\rm att}=QH_0Q^{-1}$ on $QD(H_0)$ is self-adjoint in
$\mathcal H_{\rm att}$. Its metric form is that of $Q^{-2}$, with
form domain $\operatorname{Ran}Q$. Although this range is dense in the
reference space, it is not closed in the reference norm; the transported
norm supplies its Hilbert-space structure. The generalized family
$\boldsymbol\Phi_s=Q\boldsymbol\Upsilon_s$ has the same spectral
normalization in this norm as the reference Mellin family.

\subsection{Dressing and the physical Hilbert space}

The entries $\widehat c_\pm=2/(1+e^{\alpha_\pm r})$ of $C$ are bounded,
continuous, and nonzero for finite $r\ge0$. Indeed,
$\Re\alpha_\pm=1/\sqrt2>0$, so $e^{\alpha_\pm r}\ne-1$ for $r>0$;
also $\widehat c_\pm(0)=1$ and
$\widehat c_\pm(r)\sim2e^{-\alpha_\pm r}$ at infinity.
Thus $C$ is bounded and injective. Its adjoint is injective as well,
which implies dense range. The inverse multiplication operator is
unbounded and has domain
\begin{equation}
 D(C^{-1})=\{g\in\mathcal H_0:C^{-1}g\in\mathcal H_0\}.
\end{equation}
It follows that $S=CQ$ is bounded and injective. Its range is dense
because $S^\dagger=QC^\dagger$ is injective. The inverse
\begin{equation}
 T=S^{-1}=Q^{-1}C^{-1},\qquad D(T)=\operatorname{Ran}S
 \label{eq:app-T-domain}
\end{equation}
is closed and densely defined in $\mathcal H_0$.

Define, as a set, $\mathcal H_{\rm phys}=\operatorname{Ran}S$ and equip
it with
\begin{equation}
 \langle Sf,Sg\rangle_{\rm phys}
 =\langle f,g\rangle_{\mathcal H_0}
 =\langle T(Sf),T(Sg)\rangle_{\mathcal H_0}.
 \label{eq:app-physical-product}
\end{equation}
Injectivity makes this definition unambiguous. If $Sf_n$ is Cauchy in
the physical norm, $f_n$ is Cauchy in $\mathcal H_0$, converges to $f$,
and $Sf_n$ converges to $Sf$ in the physical norm. Therefore
$\mathcal H_{\rm phys}$ is complete, and
$S:\mathcal H_0\to\mathcal H_{\rm phys}$ is unitary. In contrast,
$S$ is not unitary as a map from $\mathcal H_0$ into itself.
The embedding is continuous:
\begin{equation}
 \|\psi\|_{\mathcal H_0}\le\|S\|\,\|\psi\|_{\rm phys}.
\end{equation}
Completeness in the physical norm does not require that
$\operatorname{Ran}S$ be closed in the reference norm.

Since $T$ is closed and densely defined, $\eta=T^\dagger T$ is a
positive self-adjoint operator in the reference space. Its closed
quadratic form has domain
\begin{equation}
 D(\eta^{1/2})=D(T)=\mathcal H_{\rm phys},\qquad
 \langle\psi,\varphi\rangle_{\rm phys}
 =\langle T\psi,T\varphi\rangle_{\mathcal H_0}.
 \label{eq:app-metric-form}
\end{equation}
The notation $\langle\psi,\eta\varphi\rangle$ additionally requires
$\varphi\in D(\eta)$. This form formulation avoids assigning an
unbounded metric to vectors outside its domain. Domain distinctions
are essential in pseudo-Hermitian constructions with unbounded
metrics~\cite{MostafazadehUnbounded}.

\subsection{Self-adjoint dynamics and the real spectral line}

The logarithmic-coordinate unitary map
\begin{equation}
 (Uf)(x)=e^{x/2}f(e^x),\qquad x\in\mathbb R,
\end{equation}
transforms $H_0=-i\sigma^z A$ into $-i\sigma^z\partial_x$.
The operator on compactly supported smooth spinors is essentially
self-adjoint; its closure has domain
$U^{-1}[H^1(\mathbb R)\otimes\mathbb C^2]$ and purely absolutely
continuous spectrum $\mathbb R$ of multiplicity two. A normalized
channel basis is
\begin{equation}
 \begin{aligned}
 u_{t,+}(r)&=\frac1{\sqrt{2\pi}}
 \begin{pmatrix}r^{-1/2+it}\\0\end{pmatrix},\\
 u_{t,-}(r)&=\frac1{\sqrt{2\pi}}
 \begin{pmatrix}0\\r^{-1/2-it}\end{pmatrix}.
 \end{aligned}
 \label{eq:app-Mellin-channels}
\end{equation}
Both channels have energy $t$.

The unitary transport defines
\begin{equation}
 H_{\rm phys}=SH_0S^{-1},\qquad D(H_{\rm phys})=SD(H_0).
 \label{eq:app-transported-H}
\end{equation}
It is self-adjoint in $\mathcal H_{\rm phys}$ and has the same spectrum
and multiplicity as $H_0$. On the transported core,
Eq.~\eqref{eq:app-heat-intertwining} identifies its differential
expression with $C[-i\sigma^z(A+B)]C^{-1}$. Symmetry in the physical
product can be written without ambiguous adjoint domains as
\begin{equation}
 \langle H_{\rm phys}\psi,\varphi\rangle_{\rm phys}
 =\langle\psi,H_{\rm phys}\varphi\rangle_{\rm phys},
 \quad\psi,\varphi\in D(H_{\rm phys}).
\end{equation}
The conventional identity $H^\dagger\eta=\eta H$ expresses this
pseudo-Hermiticity wherever the reference-space compositions are
defined; it must not be read as an unrestricted identity on
$\mathcal H_0$.

\subsection{Integral spinors as generalized Mellin states}

Setting $u=p/x$ and $v=px$ in Eq.~\eqref{eq:Is-definition} gives
\begin{align}
 I_s(r)&=b_s\int_0^\infty x^{-s}e^{-\pi x^2}\sin(rx)\,dx,
 \label{eq:app-Is-sine}\\
 b_s&=\pi^{-(s+1)/2}\Gamma\!\left(\frac{s+1}{2}\right).
 \nonumber
\end{align}
For $0<\Re s<1$, removing the Gaussian in the generalized
sine-transform representation gives
\begin{align}
 Q^{-1}I_s(r)&=a_s r^{s-1},\nonumber\\
 a_s&=b_s\Gamma(1-s)\cos(\pi s/2)=(2\sqrt\pi)^{-s}\Gamma(1-s/2).
 \label{eq:app-a-coefficient}
\end{align}
Equivalently, forward heat evolution maps $a_s r^{s-1}$ to $I_s$;
this defines the inverse operation on this generalized family.
It follows that
\begin{equation}
 \boldsymbol\Upsilon_s=
 \begin{pmatrix}a_s r^{s-1}\\a_{1-s}r^{-s}\end{pmatrix},\quad
 H_0\boldsymbol\Upsilon_s=-i(s-\tfrac12)\boldsymbol\Upsilon_s.
 \label{eq:app-Upsilon}
\end{equation}
For this to be a state of the real spectral representation of the
self-adjoint $H_0$, one must have $s=1/2+it$. Off-line powers still
solve the differential expression, but they are excluded as spectral
states of this realization. No zero-selection condition is needed
for the reality of its spectrum.

Write $a_t=a_{1/2+it}$ and denote the associated spinors by
$\boldsymbol\Upsilon_t$ and $\boldsymbol\Psi_t$. Since
$\int_0^\infty r^{-1+i\omega}dr=2\pi\delta(\omega)$,
\begin{equation}
 \langle\boldsymbol\Upsilon_t,\boldsymbol\Upsilon_{t'}\rangle
 =4\pi|a_t|^2\delta(t-t').
 \label{eq:app-Upsilon-delta}
\end{equation}
Thus $u_t=\boldsymbol\Upsilon_t/(\sqrt{4\pi}|a_t|)$ is delta normalized,
and its transported state has the same normalization in the physical
metric. Individual $\boldsymbol\Psi_t$ are generalized states of the
physical space, not finite-norm vectors in that space, even though
the dressing makes them square integrable in the reference norm.
Their logarithmically divergent individual metric norm is analogous
to that of plane waves and does not invalidate the spectral construction.

For wave packets in this channel,
\begin{equation}
 f=\int_{\mathbb R}c(t)u_t\,dt,\qquad
 \psi=Sf,\qquad
 \|\psi\|_{\rm phys}^2=\int_{\mathbb R}|c(t)|^2dt,
 \label{eq:app-wave-packets}
\end{equation}
with $c\in L^2(\mathbb R,dt)$. All uses of the inverse metric
transformation on individual states are understood through this
spectral extension, rather than by applying $T$ to them as ordinary
vectors in $D(T)$.

\subsection{$PT$ symmetry and the integral-family channel}

The coefficients obey $a_{\bar s}=\overline{a_s}$, so on the critical
line $a_{1-s}=\overline{a_s}$. Hence
\begin{equation}
 \Theta\boldsymbol\Upsilon_t=\boldsymbol\Upsilon_t,
 \qquad\Theta=\sigma^x\mathcal K.
\end{equation}
The real scalar heat kernel and
$\sigma^x\overline C\sigma^x=C$ imply that $S\Theta=\Theta S$.
The transported domain is $PT$ invariant, and
$\Theta\boldsymbol\Psi_t=\boldsymbol\Psi_t$.

At fixed energy $t$, an arbitrary channel has amplitudes $(u,v)$
in the basis \eqref{eq:app-Mellin-channels}. The condition of being
fixed by $PT$ is $v=\bar u$, which is real-linear and does not alone
reduce the complex spectral multiplicity. The integral family supplies
the stronger complex-linear channel relation
\begin{equation}
 v=\rho(t)u,\qquad \rho(t)=\frac{\overline{a_t}}{a_t},\qquad
 |\rho(t)|=1.
 \label{eq:app-one-channel}
\end{equation}
Its normalized vectors span a closed reducing subspace in the
spectral representation, isomorphic to $L^2(\mathbb R,dt)$. The
restriction of $H_0$ to this subspace, and its transport by $S$, have
multiplicity one. The subspace is $PT$ invariant; individual complex
multiples need not be fixed by $PT$. The relative phases in
Eq.~\eqref{5.12} specify the cancellation of its dressed amplitudes.

\subsection{Selection of points from the continuum}

Within the self-adjoint realization, Eq.~\eqref{5.12} is evaluated as
$R(1/2+it)=0$. It selects isolated real energies, since $R$ is analytic
and not identically zero. Reality is inherited from the dilation
representation, rather than produced by this condition.

A discrete set of selected generalized states is not automatically
point spectrum of a new self-adjoint operator. For a discrete set
$Z\subset\mathbb R$, the absolutely continuous spectral measure gives
$P_{H_0}(Z)=0$. Thus such states do not form a nontrivial subspace of
normalizable packets supported on $Z$ in the existing $dt$ measure.
A discrete Hilbert-space realization, if desired, requires an additional
construction and justification; Dirac-delta normalization cannot simply
be replaced by Kronecker normalization.

Finally, the physical spectral problem and the analytic zero problem
must be distinguished. Self-adjointness excludes complex energies from
the physical spectrum. To deduce RRH for the analytic function, one
must additionally show that every zero of $R(s)$ in the relevant strip
is represented by an admissible state of this physical realization.
The heat transformation establishes the real continuum and its metric;
this remaining correspondence is not implied by merely solving the
local differential equations for arbitrary complex $s$.


\section{Exponential arithmetic filtering, prime dilations, and periodic-orbit structure}
\label{app:arithmetic-prime-filter}

\subsection{The logarithmic generator and its domain}
The dressing coefficients in Eq.~\eqref{5.9} have the continuous logarithm
\begin{equation}
 -\log\widehat c_\pm(r)=v_\pm(r)
 =\alpha_\pm r-\log2+\log(1+e^{-\alpha_\pm r}).
 \label{eq:app-arithmetic-v}
\end{equation}
For $r>0$, $|e^{-\alpha_\pm r}|<1$, fixing the final logarithm by its
convergent series. This defines $v_\pm(0)=0$ and avoids phase jumps of
a pointwise principal logarithm of $\widehat c_\pm$. The multiplication
operator $V_{\rm ar}=\operatorname{diag}(v_+,v_-)$ has maximal domain
\begin{equation}
 D(V_{\rm ar})=\{f\in\mathcal H_0:V_{\rm ar}f\in\mathcal H_0\},
 \qquad C=e^{-V_{\rm ar}}.
\end{equation}
This is analogous to $Q=e^{-\pi K}$ as an exponential filter, but its
generator is a complex multiplication profile rather than a Laplacian.

\subsection{A convergent operator Euler product}
On the scalar space $L^2((0,\infty),dr)$, define
\begin{equation}
 (\mathsf D_n f)(r)=f(nr),\qquad
 \mathsf D_m\mathsf D_n=\mathsf D_{mn},\qquad
 \|\mathsf D_n\|=n^{-1/2}.
\end{equation}
The operator series
\begin{equation}
 \mathcal Z=\sum_{n\geq1}\frac{\mathsf D_n}{n}
 =\prod_p\left(I-\frac{\mathsf D_p}{p}\right)^{-1}
 \label{eq:app-prime-Euler-product}
\end{equation}
converges in norm, since $\sum_n n^{-3/2}<\infty$. Each factor expands
in a norm-convergent geometric series; unique factorization identifies
the limit of finite prime products with the series over all integers.

For the seed $f_\alpha(r)=e^{-\alpha r}$, with $\Re\alpha>0$,
\begin{align}
 (\mathcal Zf_\alpha)(r)&=-\log(1-e^{-\alpha r}),\\
 [(I-\mathsf D_2)\mathcal Zf_\alpha](r)
 &=\log(1+e^{-\alpha r}).
 \label{eq:app-prime-seed}
\end{align}
These identities hold pointwise for $r>0$ and in $L^2$. Inserting
$\alpha=\alpha_\pm$ proves Eq.~\eqref{eq:main-prime-profile}.
The linear and constant terms of $v_\pm$ are not $L^2$ vectors;
together with the seed-generated function they define a multiplication
profile. The prime product constructs this profile and is not itself
an Euler product of independent multiplication filters $C_p$.

\subsection{Prime powers and logarithmic translations}
The logarithm of the operator product is
\begin{equation}
 \mathcal P=\sum_p\sum_{m\geq1}\frac{\mathsf D_{p^m}}{m p^m},
 \qquad \mathcal Z=e^{\mathcal P}.
\end{equation}
The sum is absolutely convergent in norm because
$\sum_{p,m}p^{-3m/2}/m<\infty$. For the scalar self-adjoint dilation
generator $h=-i(r\partial_r+1/2)$,
\begin{equation}
 \mathsf D_n=n^{-1/2}e^{i(\log n)h},\qquad
 \mathcal Z=\zeta(\tfrac32-ih).
 \label{eq:app-prime-spectral-calculus}
\end{equation}
The domain of $h$ is $U^{-1}H^1(\mathbb R)$ with
$(Uf)(x)=e^{x/2}f(e^x)$. Thus $Ue^{iah}U^{-1}g(x)=g(x+a)$:
prime powers supply translations of length $m\log p$.
The zeta function in Eq.~\eqref{eq:app-prime-spectral-calculus} is
evaluated entirely in its absolutely convergent half-plane; the
identity neither assumes RH nor directly samples its critical-line zeros.

There are two distinct exponential operations in this construction:
$e^{\mathcal P}$ acts on a seed to produce a function, whereas
$e^{-V_{\rm ar}}$ multiplies a state by the resulting filter.
Confusing these exponentials would incorrectly identify $C$ with
$\zeta(3/2-ih)$.

\subsection{The von Mangoldt operator}
\label{app:von-mangoldt-operator}

The prime-power logarithm can be written using the von Mangoldt
function $\Lambda(n)$, which equals $\log p$ for $n=p^m$ with $p$
prime and $m\geq1$, and vanishes otherwise. Introduce the real
parameter $a>1/2$ and set
\begin{equation}
 \mathcal Z_a=\prod_p(I-p^{-a}\mathsf D_p)^{-1}
 =e^{\mathcal P_a},\qquad
 \mathcal P_a=\sum_{n\geq2}
   \frac{\Lambda(n)}{n^a\log n}\,\mathsf D_n .
 \label{eq:app-vm-euler}
\end{equation}
For $n=p^m$, the coefficient $\Lambda(n)/\log n$ is $1/m$,
so this is the logarithm defined by the convergent Euler-product
series. At $a=1$ it gives $\mathcal Z_1=\mathcal Z$ and
$\mathcal P_1=\mathcal P$.

The operator series and its termwise derivative converge absolutely
in norm, locally uniformly for $a>1/2$. Indeed,
$\|\mathsf D_n\|=n^{-1/2}$ and $\Lambda(n)\leq\log n$, so on any
$a\geq a_0>1/2$ the differentiated series is bounded in norm by the
convergent sum $\sum_{n\geq2}(\log n)n^{-a_0-1/2}$. Thus the
operator logarithmic derivative is
\begin{equation}
 \mathcal M_a:=-\partial_a\mathcal P_a
 =\sum_{n\geq2}\frac{\Lambda(n)}{n^a}\,\mathsf D_n
 =\sum_{n\geq2}\frac{\Lambda(n)}{n^{a+1/2}}\,
    e^{i(\log n)h}.
 \label{eq:app-vm-operator}
\end{equation}
Differentiation supplies a factor $\log(p^m)=m\log p$ at each
prime power, leaving the weight $\log p$. In the unitary Mellin
representation, with transform kernel
$(2\pi)^{-1/2}r^{-1/2-it}$, $h$ acts as multiplication by $t$.
Equation~\eqref{eq:app-vm-operator} therefore has the symbol
\begin{equation}
 -\frac{\zeta'(a+1/2-it)}{\zeta(a+1/2-it)}.
 \label{eq:app-vm-symbol}
\end{equation}
This is the Dirichlet series for the ordinary zeta logarithmic
derivative in its half-plane of absolute convergence
\cite{Titchmarsh86}. The critical line would correspond to $a=0$,
outside the range of the norm-convergent construction.

The operator gives a direct pairing with weighted prime-power sums.
Writing $g=Uf$ for a reference packet and using the translation
action of $e^{i(\log n)h}$, one obtains
\begin{equation}
 \langle f,\mathcal M_a f\rangle_{L^2(dr)}
 =\sum_{n\geq2}\frac{\Lambda(n)}{n^{a+1/2}}\,
    \langle g,g(\,\cdot+\log n)\rangle_{L^2(dx)} .
 \label{eq:app-vm-packet}
\end{equation}
The same arithmetic coefficients enter the Guinand--Weil explicit
formula and the weighted von Mangoldt sums discussed in
Alp\"oge's recent announcement of the proof of the RH ~\cite{AlpogeRemark2026}. A comparison with
the prime side of that formula would require specifying a test
functional, accounting for the inverse translations and the
archimedean terms, and identifying the corresponding sum over zeros.
In particular, the Mellin multiplier~\eqref{eq:app-vm-symbol}
alone supplies neither a trace identity for $H_{\rm phys}$ nor
a positivity criterion for the zeros of $\zeta_{\rm reg}$.

\subsection{Comparison with periodic-orbit expansions}
On a generalized scalar dilation eigenstate with energy $E$,
\begin{equation}
 \mathcal P(E)=\sum_p\sum_{m\geq1}
 \frac{e^{-\frac32mT_p}}m e^{imET_p}
 =\log\zeta(\tfrac32-iE),\qquad T_p=\log p.
 \label{eq:app-prime-orbit-analogy}
\end{equation}
This has the algebraic structure of a logarithmic expansion over
primitive objects $p$ and their repetitions $m$. The phase corresponds
formally to an action $S_{p,m}(E)=mE\log p$, with
$\partial_E S_{p,m}=m\log p$ in units $\hbar=1$.
Such prime periods are central to the periodic-orbit analogy for
Riemann zeros~\cite{BK99}; the Dirac--Rindler mirror construction also
provides a concrete setting for prime-related returns~\cite{GS16}.

In the present representation, however, translation on the full
logarithmic line is not a closed orbit. A return map, reflection
mechanism, or another enlarged dynamics has not been specified.
Moreover, the factor $p^{-3m/2}$ follows exactly from the arithmetic
weight $p^{-m}$ and the $L^2(dr)$ normalization $p^{-m/2}$ of dilations.
It is not yet a derived Lyapunov stability factor. The series is an operator/profile construction, not a trace of a
trace-class propagator or a spectral determinant of $H_{\rm phys}$.
A periodic-orbit realization would require establishing these missing
links and its relation to the integrated zero-selection condition.

\end{document}